\documentclass[lettersize,journal]{IEEEtran}

\usepackage{amsmath,amsfonts}
\usepackage{array}
\usepackage{textcomp}
\usepackage{stfloats}
\usepackage{balance}
\usepackage{verbatim}
\usepackage[dvipsnames]{xcolor}
\usepackage{cite}
\usepackage{tikz}
\usepackage{listings}
\usepackage{siunitx}
\usepackage{xspace}
\usepackage{threeparttable}
\usepackage{multirow}
\usepackage{makecell}
\usepackage{booktabs}
\usepackage{longtable}
\usepackage{graphicx}
\usepackage{algorithmic}
\usepackage[caption=false,font=normalsize,labelfont=sf,textfont=sf]{subfig}
\usepackage{url}
\usepackage{xurl}  
\usepackage{microtype}
\usepackage{pifont} 
\usepackage{hyperref}
\hypersetup{
  colorlinks,
  linkcolor={green!80!black},
  citecolor={red!70!black},
  urlcolor={blue!70!black}
}

\ifdefined\bodyonly
  \input{body_extrefs}
  \newcommand{\extref}[2]{#2~\csname extlabel@#1\endcsname}
\else
  \newcommand{\extref}[2]{\hyperref[#1]{#2~\ref*{#1}}}
\fi
\newcommand{\apref}[1]{\extref{#1}{Appendix}}

\newcommand{\tool}[1]{#1}

\newcommand{\sysname}{\textsf{FuzzAgent}\xspace}

\lstdefinelanguage{FuzzAgent}{
  sensitive = true,
  keywords={call, load, update, assert, file},
  morekeywords=[2]={mut*, const*},
  comment=[l]{//},
  morestring=[b]',
  morestring=[b]"
}

\begin{document}
\bstctlcite{IEEEexample:BSTcontrol}

\title{FuzzAgent: Multi-Agent System for End-to-End Evolutionary Library Fuzzing}

\author{Yunlong Lyu,
        Peng Chen,
        Fengyi Wu,
        Junzhe Yu,
        and Hao Chen,~\IEEEmembership{Fellow,~IEEE}
\thanks{Y. Lyu, J. Yu, and H. Chen are with The University of Hong Kong, Hong Kong, Hong Kong (e-mail: yunlong.lyu97@gmail.com; junzheyu1@connect.hku.hk; chenho@hku.hk).}
\thanks{P. Chen is an independent researcher (e-mail: spinpx@gmail.com).}
\thanks{F. Wu is with Southeast University, Nanjing, Jiangsu, China (e-mail: fengyiwucs@gmail.com).}
\thanks{Manuscript received XXXX XX, 2026; revised XXXX XX, 2026.}}

\markboth{IEEE Transactions on XXXX,~Vol.~XX, No.~X, XXXX~2026}%
{Lyu \MakeLowercase{\textit{et al.}}: FuzzAgent: Multi-Agent System for End-to-End Evolutionary Library Fuzzing}

\maketitle

\begin{abstract}

Library fuzzing synthesizes harnesses that expose new API interactions, but deploying it end to end still requires extensive project-specific integration. Existing LLM-based systems automate individual steps, yet they treat execution outcomes as coarse signals: a stalled campaign does not reveal which state blocks progress, and a crash trace does not reveal whether the library or the harness is at fault. We present \sysname, a multi-agent system that builds and evolves a complete fuzzing campaign from a source tree alone. Specialized agents acquire project knowledge and repair campaign artifacts automatically. Hierarchical coverage diagnosis localizes blockers from modules down to branches, and dependency-aware generation converts them into harnesses with the required API lifecycle. Crash validation recovers runtime object states to distinguish harness misuse from library bugs. 

We evaluate \sysname on 20 real-world C/C++ libraries against five state-of-the-art baselines (\tool{OSS-Fuzz}, \tool{OSS-Fuzz-Gen}, \tool{PromptFuzz}, \tool{PromeFuzz}, and \tool{ProteusFuzz}). \sysname completes the full fuzzing lifecycle for all 20 libraries without human intervention at \$3.17 per library. On the 20 libraries, it reaches \num{179619} branches, exceeding \tool{OSS-Fuzz}, \tool{PromptFuzz}, \tool{ProteusFuzz}, \tool{PromeFuzz}, and \tool{OSS-Fuzz-Gen} by 45.1\%, 73.2\%, 87.7\%, 92.1\%, and 191.2\%, respectively. \sysname also identifies 102 new library bugs, 90 of which have already been acknowledged or fixed by upstream maintainers.

\end{abstract}

\begin{IEEEkeywords}
Library fuzzing, large language models, multi-agent systems, vulnerability detection.
\end{IEEEkeywords}

\section{Introduction}

Fuzzing is widely adopted to repeatedly execute programs with generated inputs to expose crashes, memory errors, and other vulnerabilities~\cite{tase_2021_fuzzing_survey,fuzzers_for_stateful_systems,fuzzing_vulnerability_discovery_survey}. Advances in coverage guidance, structured input generation, and program analysis have substantially improved its ability to explore complex software~\cite{afl,aflfast,sec_2023_dynsql,sec_2020_symcc}. As of 2026, the continuous fuzzing platform \tool{OSS-Fuzz}~\cite{oss_fuzz} has reported more than \num{50000} bugs in open-source software. However, this success depends on extensive project-specific integration where experts must prepare build scripts, harnesses, dictionaries, and seed corpora for each target. For library targets, further exploration is constrained not only by input generation but also by these harnesses~\cite{fuzzing_2023_fuzz_blockers}. This observation motivates library fuzzing, which treats the harness itself as a search object~\cite{fudge,fuzzgen,apicraft,graphfuzz,hopper,ccs_24_promptfuzz,ccs_25_promefuzz}. By synthesizing drivers that map fuzzer-generated bytes to API arguments and coordinate sequences of library calls, library fuzzing can construct new program states and expose behaviors omitted from manually integrated targets.

Automated library-fuzzing systems commonly infer API usage knowledge, generate candidate harnesses, validate their execution, and use the resulting coverage to guide further generation. Representative systems implement this workflow with progressively richer forms of program knowledge. \tool{Hopper}~\cite{hopper} uses program analysis and predefined rules to infer API constraints and construct executable call sequences. \tool{PromptFuzz}~\cite{ccs_24_promptfuzz} forms prompts from mutated API combinations, asks an LLM to synthesize harnesses, and filters the candidates through compilation, execution, and short fuzzing. \tool{PromeFuzz}~\cite{ccs_25_promefuzz} further preconstructs codebase summaries and retrieves API-related context to support harness generation and one-step LLM-based crash triage. \tool{ProteusFuzz}~\cite{tdsc_26_proteusfuzz} extends \tool{PromeFuzz} with an agent that is activated when coverage plateaus, diagnoses the blocking condition, and revises seeds or harnesses accordingly. Although their mechanisms differ, they exemplify a common paradigm: statically assembled knowledge drives harness generation, and execution outcomes steer exploration through coarse signals or a stagnation-triggered repair step.

This paradigm leaves three system-level challenges, as detailed in \autoref{sec:challenges}. First, harness synthesis is only one stage of an interdependent fuzzing lifecycle. Starting from source code alone requires operational and input-domain knowledge for building, instrumentation, auxiliary inputs, execution, and validation; this knowledge is scattered across source files, documentation, and external resources, while a failure in one stage may invalidate artifacts consumed downstream. Second, a harness that compiles, passes validation, and reaches an API may still exercise only shallow behavior. \tool{ProteusFuzz}~\cite{tdsc_26_proteusfuzz} shows that such blockers can be detected from coverage stagnation, but two problems remain: feedback must be fine-grained enough to guide the repair, from subsystems down to branch evidence, rather than a single stagnation line, and the revised harness needs the complete API lifecycle, not a bare call chain. Third, generated harnesses can violate implicit contracts and thereby produce crashes deep inside library code. Call traces and statically retrieved API summaries localize where such failures manifest, but they do not reveal the concrete runtime state or necessarily retrieve the non-local contract needed to distinguish harness misuse from a library defect. For effective library fuzzing, current systems still face significant challenges in automation, effective harness generation and library crash validation.

We present \sysname, an evolutionary multi-agent system that automates library fuzzing from source code alone. Specialized interfaces ground decisions in project and execution state, enabling LLM agents to iteratively refine campaign setup, harnesses, and crash verdicts. For end-to-end automation, specialized agents acquire the required project knowledge to construct lifecycle artifacts, and use interface-bounded diagnostics to execute, verify, and repair them. For coverage evolution, hierarchical diagnosis connects module, file, API, function, and branch coverage, allowing the system to convert a concrete coverage gap into an exploration blueprint, while dependency-aware generation reconstructs four types of API dependencies to realize the intended behavior. For crash validation, the system recovers concrete runtime object states and constructs adaptive contracts for grounding each verdict in the relevant rule.


We evaluate \sysname on 20 real-world C/C++ libraries against \tool{OSS-Fuzz}, \tool{OSS-Fuzz-Gen}, \tool{PromptFuzz}, \tool{PromeFuzz}, and \tool{ProteusFuzz}. Across five 24-hour end-to-end rounds, \sysname completes all rounds without human intervention, at an average LLM cost of \$3.17 per target. Across the 20 libraries, it covers \num{179619} branches, improvements of 45.1\%, 73.2\%, 87.7\%, 92.1\%, and 191.2\% over \tool{OSS-Fuzz}, \tool{PromptFuzz}, \tool{ProteusFuzz}, \tool{PromeFuzz}, and \tool{OSS-Fuzz-Gen}~\cite{oss-fuzz-gen}, respectively. Finally, \sysname identifies 102 new library bugs among 128 candidates and 90 of these bugs have been acknowledged or fixed by upstream maintainers. We release the code and experimental data of \sysname at \url{https://github.com/FuzzAnything/FuzzAgent}.


\section{Challenges}
\label{sec:challenges}

Effective library fuzzing must autonomously construct an executable campaign, turn coverage gaps into harnesses that reach new program states, and distinguish library defects from harness-induced failures. The following subsections show how these challenges limit the automation, exploration, and detection reliability of existing systems.

\definecolor{CoverageState}{RGB}{0,78,91}
\definecolor{CoverageBlocker}{RGB}{165,42,42}
\definecolor{CoverageAction}{RGB}{0,105,62}
\newcommand{\CoverageReachedPhrase}[1]{%
  \begingroup\setlength{\fboxsep}{1pt}%
  \colorbox{CoverageAction!12}{\textcolor{CoverageAction!80!black}{\strut #1}}%
  \endgroup}
\newcommand{\CoverageBlockedPhrase}[1]{%
  \begingroup\setlength{\fboxsep}{1pt}%
  \colorbox{BurntOrange!16}{\textcolor{BurntOrange!65!black}{\strut #1}}%
  \endgroup}
\providecommand*{\subfigureautorefname}{Figure}
\makeatletter
\newcommand{\HarnessSubfigureLabel}[1]{%
  \protected@edef\@currentlabel{\thefigure(\alph{subfigure})}%
  \label{#1}}
\makeatother
\begin{figure*}[!t]
  \newcommand{\CoverageCodeLineNumber}{%
    \ttfamily\scriptsize
    \ifnum\value{lstnumber}>0
      \ifnum\value{lstnumber}<5\color{black}\else\color{red!2}\fi
    \else\color{red!2}\fi}
  \centering
  \subfloat{\HarnessSubfigureLabel{fig:motivating_harness_shallow}}%
  \begin{minipage}[t]{0.31\textwidth}
  \textbf{(a) Valid but shallow harness}
  \begin{lstlisting}[language=C++,numbers=left,basicstyle=\ttfamily\scriptsize,backgroundcolor=\color{gray!3}]
z_stream s = {};
(*@\textcolor{RoyalBlue}{\ttfamily deflateInit2}@*)(&s, level, Z_DEFLATED,
    MAX_WBITS + 16, 8, strategy);

s.next_in = input;
s.next_out = output;
// s.gzhead is not set, so the default gzip header is used.

(*@\textcolor{CoverageState}{\bfseries\ttfamily deflate}@*)(&s, Z_FINISH);
// Compiles; runs; API reached.
  \end{lstlisting}
  \end{minipage}\hfill
  \subfloat{\HarnessSubfigureLabel{fig:motivating_harness_diagnosis}}%
  \begin{minipage}[t]{0.31\textwidth}
  \textbf{(b) Diagnosis feedback}\par
  \begin{lstlisting}[numbers=none,basicstyle=\ttfamily\scriptsize,backgroundcolor=\color{CoverageState!4},frame=single,rulecolor=\color{CoverageState!55},framerule=0.6pt]
(*@\textcolor{gray}{\bfseries\sffamily Stagnation line (ProteusFuzz)}@*)
  Line: s->gzhead == Z_NULL
  API:  deflate

(*@\textcolor{CoverageAction}{\bfseries\sffamily Hierarchical coverage (ours)}@*)
  File   deflate.c (*@\textcolor{BurntOrange}{\bfseries\ttfamily [partial]}@*)
  API    deflateInit2 (*@\textcolor{CoverageAction}{\bfseries\ttfamily [covered]}@*)
         deflate (*@\textcolor{CoverageAction}{\bfseries\ttfamily [covered]}@*)
         deflateSetHeader (*@\textcolor{BurntOrange}{\bfseries\ttfamily [uncovered]}@*)
  Func   deflate (*@\textcolor{BurntOrange}{\bfseries\ttfamily [partial]}@*)
  Branch s->gzhead == Z_NULL
  \end{lstlisting}
  \end{minipage}\hfill
  \subfloat{\HarnessSubfigureLabel{fig:motivating_harness_revision}}%
  \begin{minipage}[t]{0.34\textwidth}
  \textbf{(c) Dependency-aware harness revision}
  \begin{lstlisting}[language=C++,numbers=left,basicstyle=\ttfamily\scriptsize,backgroundcolor=\color{green!2}]
z_stream s = {};
(*@\textcolor{CoverageAction}{\bfseries\ttfamily deflateInit2}@*)(&s, level, Z_DEFLATED,
    (*@\textcolor{CoverageAction}{\bfseries\ttfamily MAX\_WBITS + 16}@*), 8, strategy);

gz_header h = {}; 
h.extra = extra;
h.name = name; h.comment = comment;
(*@\textcolor{CoverageAction}{\bfseries\ttfamily deflateSetHeader}@*)(&s, &h);

(*@\textcolor{CoverageAction}{\bfseries\ttfamily deflate}@*)(&s, Z_FINISH);
// Executes the blocked path.
  \end{lstlisting}
  \end{minipage}
  \caption{\small{A real \tool{zlib} campaign illustrating how feedback resolution changes harness generation. (a) A harness that passes validation and reaches \texttt{deflate}. (b) A stagnation-line diagnosis provides only sparse feedback about the blocker line, whereas hierarchical coverage adds file, API, function, and branch evidence that guides the revision in (c). (c) Relating the blocking predicate to the public API contract yields an ordered API group that changes the relevant state.}}
  \label{fig:motivating_harness}
\end{figure*}

\subsection{Challenges in Automating Library Fuzzing}
\label{sec:background:library_fuzzing}


Library fuzzing is not a single code-generation task: a usable campaign must \textit{build and instrument the target}, \textit{prepare dictionaries and seed corpora}, \textit{generate and compile harnesses}, \textit{execute fuzzers}, \textit{monitor coverage and crashes}, and \textit{validate the reported failures}~\cite{oss-fuzz-guide,ccs_25_fuzzer_usability}. 
However, end-to-end deployment remains labor-intensive~\cite{oss-fuzz-guide,soups_21_clang_libfuzzer,chi_2023_fuzzer_usability,TOSEM_2023_human_side_fuzzing,ccs_25_fuzzer_usability}.
\tool{OSS-Fuzz} illustrates this boundary. It standardizes fuzzing infrastructure but still requires manually specialized build scripts, harnesses, dictionaries, and seed corpora~\cite{oss-fuzz-guide}, whose construction involves more than simply scripting. Nourry et al.~\cite{fuzzing_build_failures} examined \num{968922} \tool{OSS-Fuzz} builds and observed \num{91240} failures, showing that human-configured pipelines remain fragile to integration failures due to complexity in managing the build environment, toolchain, dependencies, and input artifacts. Evidence from human-centered studies points to the same difficulty: \num{60}\% of surveyed developers reported that fuzzing had broken their project builds~\cite{TOSEM_2023_human_side_fuzzing}; a controlled user study found that participants struggled with compiling, instrumenting, and linking fuzz targets~\cite{soups_21_clang_libfuzzer}; and practitioners described setup as an iterative process of executing the tool, interpreting its diagnostics, and revising the configuration~\cite{ccs_25_fuzzer_usability}. Together, these requirements make end-to-end campaign setup an inherently labor-intensive coordination problem.

Existing library-fuzzing systems leave the acquisition and application of this knowledge largely outside their automation boundary. Even recent harness-generation approaches~\cite{hopper,ccs_24_promptfuzz,ccs_25_promefuzz} still require experts to prepare build scripts, resolve dependencies, configure instrumentation and harness compilation, and supply dictionaries and seed corpora. In our evaluation (\autoref{sec:automation-efficiency}), onboarding a new library into \tool{PromptFuzz}~\cite{ccs_24_promptfuzz} and \tool{PromeFuzz}~\cite{ccs_25_promefuzz} required 8 and 12 hours of manual effort on average, respectively. End-to-end automation must therefore acquire and apply heterogeneous, often incomplete project knowledge and use concrete execution outcomes to refine the campaign until it succeeds.

\begin{figure*}[!t]
  \definecolor{HarnessType}{RGB}{0,78,91}
  \definecolor{HarnessEntry}{RGB}{0,88,44}
  \centering
  \subfloat{\HarnessSubfigureLabel{fig:motivating_crash_harness}}%
  \begin{minipage}[t]{0.50\textwidth}
  \vspace{0pt}%
  \textbf{(a) Minimal pseudocode from a LLM-generated harness}
  \begin{lstlisting}[language=C++,numbers=left,basicstyle=\ttfamily\scriptsize]
int (*@\textcolor{HarnessEntry}{\ttfamily LLVMFuzzerTestOneInput}@*)(const (*@\textcolor{RoyalBlue}{\ttfamily uint8\_t}@*) *data, (*@\textcolor{RoyalBlue}{\ttfamily size\_t}@*) size) {

  (*@\textcolor{RoyalBlue}{\ttfamily FuzzedDataProvider}@*) fdp(data, size);
  (*@\textcolor{RoyalBlue}{\ttfamily aom\_codec\_enc\_cfg\_t}@*) cfg;
  ...
  // Choose format and dimensions from the input.
  (*@\textcolor{RoyalBlue}{\ttfamily aom\_img\_fmt\_t}@*) fmt = fdp.(*@\textcolor{RoyalBlue}{\ttfamily ConsumeBool}@*)()
      ? (*@\textcolor{BurntOrange}{\ttfamily AOM\_IMG\_FMT\_I420}@*) : (*@\textcolor{BurntOrange}{\ttfamily AOM\_IMG\_FMT\_I444}@*);
  cfg.g_w = 2 * fdp.(*@\textcolor{RoyalBlue}{\ttfamily ConsumeIntegralInRange}@*)(8u, 512u);
  cfg.g_h = 2 * fdp.(*@\textcolor{RoyalBlue}{\ttfamily ConsumeIntegralInRange}@*)(8u, 512u);

  (*@\textcolor{RoyalBlue}{\ttfamily aom\_image\_t}@*) img = {};
  img.fmt = fmt;
  img.w = img.d_w = cfg.g_w;
  img.h = img.d_h = cfg.g_h;

  // Using the I420 layout; stride/shift omitted.
  img.planes[0] = malloc(cfg.g_w * cfg.g_h);
  img.planes[1] = malloc(cfg.g_w * cfg.g_h / 4);
  img.planes[2] = malloc(cfg.g_w * cfg.g_h / 4);

  (*@\textcolor{RoyalBlue}{\ttfamily aom\_codec\_encode}@*)(&ctx, &img, 0, 1, 0); // crashing call!
  return 0;
}
  \end{lstlisting}
  \end{minipage}\hfill
  \subfloat{\HarnessSubfigureLabel{fig:motivating_crash_trace}}%
  \begin{minipage}[t]{0.47\textwidth}
  \vspace{0pt}%
  \textbf{(b) AddressSanitizer call trace}
  \begin{lstlisting}[numbers=none,basicstyle=\ttfamily\scriptsize,backgroundcolor=\color{red!2}]
(*@\textcolor{BrickRed}{\bfseries\ttfamily heap-buffer-overflow: READ of size 475}@*)
#0 __asan_memcpy
#1 (*@\textcolor{BrickRed}{\bfseries\ttfamily copy\_and\_extend\_plane}@*) (*@\textcolor{gray}{\ttfamily av1/encoder/extend.c:36:7}@*)
#2 av1_copy_and_extend_frame (*@\textcolor{gray}{\ttfamily av1/encoder/extend.c:154:5}@*)
#3 av1_lookahead_push (*@\textcolor{gray}{\ttfamily av1/encoder/lookahead.c:151:3}@*)
#4 av1_receive_raw_frame (*@\textcolor{gray}{\ttfamily av1/encoder/encoder.c:4632:7}@*)
#5 encoder_encode (*@\textcolor{gray}{\ttfamily av1/av1\_cx\_iface.c:3418:11}@*)
#6 (*@\textcolor{RoyalBlue}{\bfseries\ttfamily aom\_codec\_encode}@*) (*@\textcolor{gray}{\ttfamily aom/src/aom\_encoder.c:191:11}@*)
#7 (*@\textcolor{ForestGreen}{\bfseries\ttfamily LLVMFuzzerTestOneInput}@*)
(*@\textcolor{gray}{\ttfamily synthesized\_driver.c:157:13}@*)
  \end{lstlisting}

  \vspace{0.4em}
  \subfloat{\HarnessSubfigureLabel{fig:motivating_crash_context}}%
  \textbf{(c) Illustrative trace-guided static context}
  \begin{lstlisting}[numbers=none,basicstyle=\ttfamily\scriptsize,backgroundcolor=\color{RoyalBlue!3}]
Retrieved API: (*@\textcolor{RoyalBlue}{\bfseries\ttfamily aom\_codec\_encode}@*)

(*@\textcolor{RoyalBlue}{\bfseries\ttfamily Summary:}@*) Encode img with an initialized
encoder; img == NULL flushes held frames.

(*@\textcolor{ForestGreen}{\bfseries\ttfamily Documented constraints:}@*)
- pts is strictly increasing.
- img != NULL implies duration > 0.
- img->fmt is supported by the encoder.
- samples fit within img->bit_depth.
  \end{lstlisting}
  \end{minipage}
  \caption{A misclassified case of \tool{PromeFuzz} on \texttt{libaom}, and its sanitizer call trace, and an illustrative context synthesized by \tool{PromeFuzz} for utilizing LLMs to triage the crash.}
  \label{fig:motivating_crash}
\end{figure*}

\subsection{Challenges in Effective Harness Generation}
\label{sec:background:harness_generation_example}

A harness is a small driver that maps fuzzer bytes to API arguments and runs a sequence of library calls. The paths it can reach depend on the APIs it calls, their order, and how input bytes map to arguments, so a useful harness must build library states that expose real behavior, not merely compile and call a target API. \tool{PromptFuzz}~\cite{ccs_24_promptfuzz} and \tool{PromeFuzz}~\cite{ccs_25_promefuzz} guide generation by harness validity and API coverage. These signals show whether an API is reached, but not what blocks the paths behind it. Such blockers are a known obstacle to fuzzing~\cite{fuzzing_2023_fuzz_blockers}, and \tool{ProteusFuzz}~\cite{tdsc_26_proteusfuzz} recently addressed them in a simple agentic loop: when coverage plateaus, an agent picks one stagnation target with a control-flow heuristic and, for state-related blockers, looks up an API call chain in a precomputed dependency graph, which an LLM uses to revise the harness. Two problems remain. We illustrate them with a real \tool{zlib} campaign (\autoref{fig:motivating_harness}) and our evaluation.

\textbf{Challenge 1: Coverage feedback is too coarse to guide blocker resolution.} The harness in \autoref{fig:motivating_harness_shallow} initializes a stream and calls \texttt{deflate}. It compiles, runs, and reaches \texttt{deflate}, so validation and API coverage both report success. However, it never sets the optional \texttt{gz\_header}, so execution always takes the \texttt{s->gzhead == Z\_NULL} branch and skips the code that handles extra data. Mutating the input cannot change this, because the state is fixed by the harness's API sequence, not by the input bytes. The API coverage used by \tool{PromptFuzz}~\cite{ccs_24_promptfuzz} and \tool{PromeFuzz}~\cite{ccs_25_promefuzz} cannot reveal it, since \texttt{deflate} counts as covered. \tool{ProteusFuzz}~\cite{tdsc_26_proteusfuzz} gives its LLM a stagnation line and the call chain from an API to that line (\autoref{fig:motivating_harness_diagnosis}). This shows where execution stops, but cannot directly guide the resolution. The LLM must still infer which state blocks execution and which other APIs can change it. Effective guidance needs coverage at each level, from modules and files down to functions and branches, and then aggregate these findings to the APIs that can resolve the blocker, such as the uncovered \texttt{deflateSetHeader} in \autoref{fig:motivating_harness_diagnosis}.

\textbf{Challenge 2: Crossing a blocker requires a dependency-aware API lifecycle.} To set a valid \texttt{gzhead}, the harness must call \texttt{deflateInit2} (configured for gzip with \texttt{MAX\_WBITS + 16}), then \texttt{deflateSetHeader}, then \texttt{deflate} (\autoref{fig:motivating_harness_revision}). \tool{ProteusFuzz}~\cite{tdsc_26_proteusfuzz} passes its LLM only the call chain from the API \texttt{deflate} to the blocked code. The other required APIs and the relations that make the sequence valid are left to the LLM to guess. The header must be built before it is passed in (construction). The stream from \texttt{deflateInit2} must be the one used by the later calls (data flow). \texttt{deflateSetHeader} must precede the first \texttt{deflate}~\cite{zlib_manual} (order). The header's buffers must stay valid until \texttt{deflate} finishes, and the stream must be released by \texttt{deflateEnd} on every path (ownership). A wrong inference either fails to cross the blocker or crosses it with an invalid sequence, and the resulting crash reflects harness misuse, not a library bug. \tool{ProteusFuzz} itself notes that revised harnesses may deviate from canonical API usage and that separating such crashes from genuine bugs is difficult~\cite{tdsc_26_proteusfuzz}. These dependencies therefore need to be recovered explicitly (see also \autoref{sec:background:crash_validation_example}).

This example exposes two related limitations of existing harness generation. First, coverage feedback is too coarse to guide blocker resolution. Second, without the dependencies of the blocked state, the generator cannot reliably build a harness that crosses it. Effective harness generation must therefore use coverage at several levels, down to branch evidence, to guide the repair, and then construct the complete API lifecycle that changes the blocked state.

\subsection{Challenges in Library Crash Validation}
\label{sec:background:crash_validation_example}
Crash reports from library fuzzing are often harder to interpret than those from standalone programs or \tool{OSS-Fuzz} integrated targets. They can be triggered by invalid API sequences or unmet preconditions in generated harnesses rather than true defects in the library implementation~\cite{fuzzing_challenges_reflections, ccs_25_fuzzer_usability}.
%
Many library fuzzing approaches~\cite{fudge, fuzzgen, intelligen, apicraft, graphfuzz, hopper, afgen, ccs_24_promptfuzz, TitanFuzz, ccs_25_promefuzz} provide limited support for library crash validation, which can lead to a high rate of false positives. Current crash filters are often based on coarse heuristics and insufficient semantic context~\cite{hopper, ccs_24_promptfuzz, ccs_25_promefuzz}. For example, \tool{Hopper}~\cite{hopper} validates crashes by predefined constraints that cannot be generalized, \tool{PromptFuzz}~\cite{ccs_24_promptfuzz} relies on short executions that may miss issues that manifest later, and \tool{PromeFuzz}~\cite{ccs_25_promefuzz} asks LLMs to judge crashes from stack traces and API summaries without concrete evidence. Consequently, \tool{PromeFuzz}, the state-of-the-art library fuzzing system, presented only 0.32\% bug detection precision in our evaluation (\autoref{sec:bug_detection})

\autoref{fig:motivating_crash} distills a representative failure pattern from a real \tool{PromeFuzz} misclassification on \tool{libaom}. These three panels show a harness that misuses an API, the associated sanitizer call trace, and the API context constructed by \tool{PromeFuzz}, respectively. In \autoref{fig:motivating_crash_harness}, the harness splits the fuzz input and chooses an image format \texttt{fmt} from two possible values, \texttt{I420} and \texttt{I444}, by consuming one fuzzed byte (line 7--8). However, the allocation of image planes uses the \texttt{I420} layout, which is incompatible with the other (line 18--20), leading to a spurious crash on the encoder API invocation when the image format \texttt{I444} is actually used (line 22).

\textbf{Challenge 1: Object states are opaque in the harness and the call trace.} At first glance, this crash exhibits several signals of a plausible library bug. In \autoref{fig:motivating_crash_harness}, the harness follows common practices to use \texttt{FuzzedDataProvider} to split the input bytes across multiple API parameters~\cite{fuzzer_split}, initializes the structures involved, and invokes the public \texttt{aom\_codec\_encode} API with typed arguments correctly.
The call trace that is deeply nested in library code in \autoref{fig:motivating_crash_trace} also reinforces this impression.  This buffer-overflow crash manifests in an internal \texttt{memcpy} after execution has passed through multiple layers of the encoder, making the failure appear to result from an incorrect bounds or address calculation deep inside the library. However, the actual cause is an inconsistency issue on image layouts that is hidden by the opacity of concrete object states in the harness.
Without inspecting runtime object state, LLMs are hard to identify the invalid execution within the static harness, while the call trace tends to reveal a bug on the validation execution path.
This opacity prevents \tool{PromeFuzz} from connecting the internal overflow to the incompatible object, leaving a library defect as a plausible but incorrect verdict.



\textbf{Challenge 2: The violated contract lies outside the static context.} \tool{PromeFuzz} does not reason from the call trace alone. It preconstructs a knowledge base from AST-extracted code metadata and implementations, documentation indexed by API name, and LLM-generated API-usage summaries~\cite{ccs_25_promefuzz}. \autoref{fig:motivating_crash_context} illustrates the resulting summary for the stack-visible \texttt{aom\_codec\_encode}: the encoder must be initialized, timestamps must increase, a non-null image requires nonzero duration, the format must be supported, and samples must fit the declared bit depth. Indeed, this summary is factually correct and the API invocation does not violate it. However, it missed the contract for defining the specification of each image layout~\cite{libaom_image_header}, which is neither an API instance nor present in the call trace.
As a result, the static context constructed by \tool{PromeFuzz} does not help LLMs to identify the actual violation.

This example exposes two related limitations of one-step LLM-based validation. First, only the harness and call trace cannot reveal the object state at runtime. Second, without that state, static retrieval cannot determine which non-local contract is relevant. Reliable validation must therefore use concrete execution state not only as evidence for the verdict, but also to construct the right context for reasoning.



\definecolor{AgentAction}{RGB}{0,78,155}
\definecolor{ToolAction}{RGB}{0,155,62}
\newcommand{\agentaction}[1]{\begingroup\setlength{\fboxsep}{0.8pt}\smash{\colorbox{AgentAction!14}{\textcolor{AgentAction!85!black}{#1}}}\vphantom{#1}\endgroup}
\newcommand{\toolaction}[1]{\begingroup\setlength{\fboxsep}{0.8pt}\smash{\colorbox{ToolAction!9}{\textcolor{ToolAction!65!black}{#1}}}\vphantom{#1}\endgroup}

\begin{figure*}[!t]
    \centering
    \includegraphics[width=0.9\linewidth]{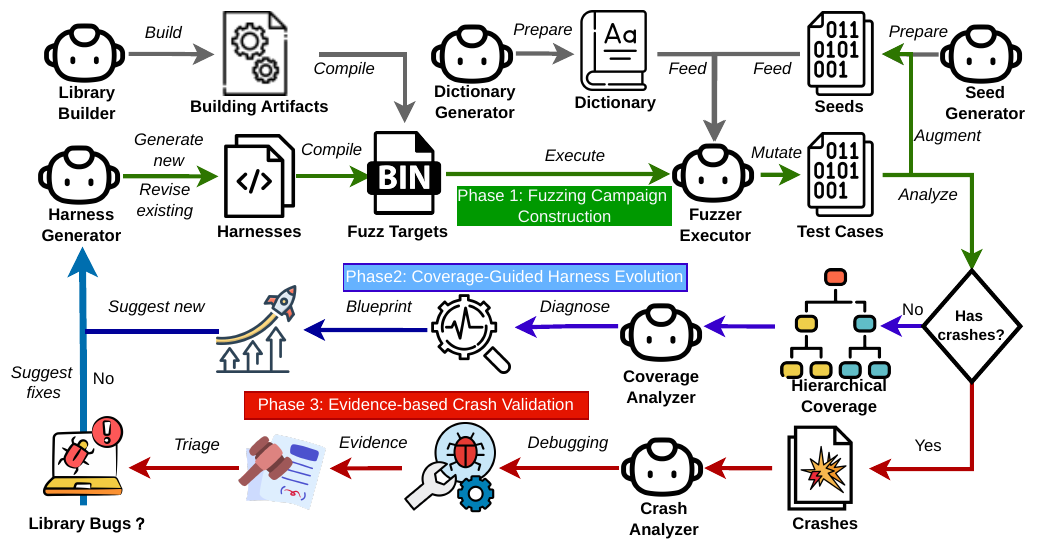}
    \caption{The workflow of evolutionary library fuzzing.}
    \label{fig:evolution}
\end{figure*}

\section{FuzzAgent: Evolutionary Library Fuzzing}
\label{sec:architecture_overview}
\label{sec:library_fuzzing_strategies}

The preceding examples expose three coupled obstacles to autonomous library fuzzing. \sysname addresses these obstacles by treating a fuzzing campaign as persistent, executable state and evolving that state from concrete evidence. Each iteration observes an outcome, diagnoses the condition responsible for it, proposes a revision, and executes the revision to validate its effect.
These evidence-guided evolutions form the methodological foundation shared by the three solutions below. Source-sufficient automation establishes and maintains an executable campaign, coverage-guided harness evolution revises that campaign to reach under-explored behavior, and evidence-based crash validation repairs semantically invalid harnesses or validates genuine library failures.

To make the division of labor explicit, key technical details below use \agentaction{by LLMs} to denote reasoning, diagnosis, and decisions performed by LLMs in agents.

\subsection{Source-sufficient Campaign Automation}
\label{sec:automation_solution}

Given only a target source code, \sysname autonomously acquires project-specific knowledge and turns it into the artifacts required to launch and sustain a complete fuzzing campaign. Specialized agents operate through domain-specific interfaces, while a stateful environment preserves artifacts and execution evidence, enabling the outcome of each stage to be validated.

\subsubsection{End-to-End Workflow}
Autonomous end-to-end library fuzzing requires coordinating the complete campaign lifecycle, from environment setup and harness execution to coverage analysis and crash validation. \sysname organizes these tasks into the three-phase workflow shown in \autoref{fig:evolution}, where each phase is conducted by dedicated agents and an agent schedules the workflow. Phase~1 constructs the campaign: the \textit{Library Builder} produces instrumented build artifacts, while the \textit{Dictionary Generator} and \textit{Seed Generator} prepare the auxiliary inputs needed for effective fuzzing. Then the \textit{Harness Generator} creates or revises a harness and the \textit{Fuzzer Executor} runs it with the prepared dictionary and seed corpus. Newly discovered test cases augment the shared corpus, so subsequent harnesses inherit the input progress of earlier campaigns. Each fuzzing campaign execution then determines the next transition. In the absence of a crash, the \textit{Coverage Analyzer} examines hierarchical coverage and diagnoses the coverage gaps to return as a concrete blueprint for enhancing the code coverage (Phase~2). If a crash occurs, the \textit{Crash Analyzer} enters Phase~3 to analyze its cause. A harness misuse yields an actionable correction and returns to harness revision, whereas a validated library fault is preserved as a bug report. 

\subsubsection{Tool-Grounded and Stateful Automation}

\noindent
\textbf{Tool-grounded domain execution.}
\sysname combines general-purpose tools with specialized interfaces (tool collections) instead of requiring agents to express every campaign operation as shell commands. File navigation, editing, and Bash retain the flexibility needed for open-ended source inspection and artifact modification. Four specialized interface families run the domain workflow: (1) the \emph{Web Interface} retrieves build recipes and input artifacts from external materials in the Web; (2) the \emph{Fuzzing Interface} builds instrumented library variants, validates dictionaries, compiles harness code, and executes fuzz targets with the campaign's corpus and configuration; (3) the \emph{Coverage Analysis Interface} extracts a hierarchical code coverage and analyzes call chains to locate coverage bottlenecks; and (4) the \emph{Crash Debugging Interface} minimizes harness code and reproduces failures, collects sanitizer and source evidence, and maintains iterative debugger sessions. These interfaces encode the invariant workflows and strategies, while leaving project-specific diagnosis and scheduling to the LLMs. This separation prevents the agents from being overwhelmed by low-level orchestration and turns their decisions into observable actions.

\noindent
\textbf{Campaign artifact setup.}
Grounded by these execution capabilities, the \textit{Library Builder} reconstructs the operational knowledge required to build the target. It identifies the build system, dependencies, and configuration options from build files, source code, and project documentation, then synthesizes a reusable build script \agentaction{by LLMs} with configurable instrumentation flags. The Fuzzing Interface executes this script to produce separate variants instrumented for fuzzing (e.g., AFL~\cite{afl}), sanitizers (e.g., ASAN~\cite{asan} and UBSAN~\cite{ubsan}), and source-based coverage~\cite{source_based_code_coverage}, and verifies both the resulting artifacts and their instrumentation. Compilation or verification failures yield concrete diagnostics for LLM-guided script revision and retry; when local evidence is insufficient, the builder consults the \textit{Web Interface} for external build knowledge. The workflow also prepares the input-domain artifacts needed to exercise the built library. The \textit{Dictionary Generator} and \textit{Seed Generator} infer the target's formats and protocols from source code and documentation \agentaction{by LLMs}, which guide the Web Interface in retrieving dictionaries and test inputs from related projects. Retrieval is particularly important for structured binary formats that current LLMs struggle to produce directly~\cite{usenix25_g2fuzz}. Moreover, seed selection materially affects fuzzing outcomes, and collecting candidates before retaining a compact, coverage-relevant corpus is more effective than starting from empty or singleton inputs~\cite{issta_21_seed_selection}. Together, the reusable instrumented builds, dictionaries, and seed corpora provide the concrete artifacts needed to launch fuzzing without manual preparation.

\noindent
\textbf{Bounded harness execution.}
After compilation, the \textit{Fuzzer Executor} invokes the Fuzzing Interface to run the harness with the prepared dictionary and shared seed corpus. Each campaign is a blocking, bounded experiment. The workflow advances only after the executor observes a crash, detects a coverage plateau, or exhausts the assigned time budget. Let $N_t$ denote the number of runtime features covered at monitoring time $t$. The executor measures relative coverage growth as
\begin{equation}
R_t = \frac{N_t-N_{t-1}}{N_{t-1}}.
\label{eq:coverage_growth_rate}
\end{equation}
A plateau is declared when $R_t$ falls below the threshold $\epsilon$ throughout a window $w$~\cite{CodaMosa}. This stopping policy prevents a saturated harness from monopolizing the end-to-end budget.

\noindent
\textbf{Stateful verification and repair.}
\sysname makes the campaign stateful by persisting every expected outcome from agents and interfaces as structured artifacts in an environment. Each agent has an artifact-level contract that specifies the artifacts it consumes and produces. Before advancing the workflow, it verifies that the required outputs exist and satisfy their postconditions. A failed build, invalid harness, or incomplete execution is therefore recorded as an intermediate state rather than accepted as a terminal result. Diagnostics are returned to responsible agents, which revise the artifact and invoke the relevant interface again. These execute--verify--repair loops prevent early-stage errors from silently propagating and stabilize the autonomous campaign.


\subsection{Coverage-Guided Harness Evolution}
\label{sec:coverage_driven_evolution}

\begin{figure*}[!t]
    \centering
    \includegraphics[width=\linewidth]{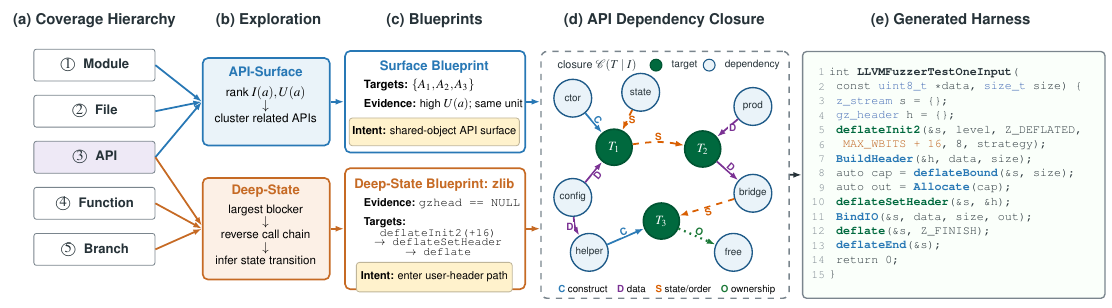}
    \caption{Coverage-guided harness evolution in \sysname.
     }
    \label{fig:coverage_guided_harness_evolution}
\end{figure*}

Coverage-guided evolution converts the current coverage state into a targeted program transformation rather than a scalar generation reward. Each iteration couples two methods: hierarchical diagnosis selects one high-value missing behavior and identifies the evidence that controls it, while dependency-aware generation constructs an input-driven API lifecycle capable of changing that state. \autoref{fig:coverage_guided_harness_evolution} summarizes this transformation from hierarchical coverage through a phase-specific blueprint and dependency closure to a generated harness. Together, these methods realize a \emph{coverage state $\rightarrow$ blocker $\rightarrow$ API group and constraints $\rightarrow$ harness generation} pipeline.

\subsubsection{Hierarchical Coverage Diagnosis}
The \textit{Coverage Analyzer} invokes the Coverage Interface to extract a coverage hierarchy that connects campaign-level progress to an exact coverage enhancement plan. The hierarchy is not a collection of isolated percentages: each level answers a different diagnostic question and supplies the context required to interpret the coverage gap. We therefore traverse it top-down, moving from coarse-grained structural localization to fine-grained semantic and control-flow evidence. This organization draws on LLVM source-based coverage metrics and extends fuzzing introspection analyses that combine runtime coverage with static call-graph reachability and code complexity to identify fuzz blockers~\cite{source_based_code_coverage,fuzz_introspector,fuzzing_2023_fuzz_blockers}.

The two upper levels provide a semantic map from library subsystems to the logical functional units implemented within them. \emph{(1) Module-level coverage} aggregates coverage along source-tree component boundaries, exposing under-explored subsystems without flattening a large library into thousands of unrelated functions. \emph{(2) File-level coverage} treats each source file as a logical functional unit within its subsystem and localizes APIs that are functionally dependent. The remaining levels connect this organizational context to concrete entry points and control flow. \emph{(3) API-level coverage} identifies public entry points, recording each API's harness invocations and accumulated uncovered branches over its reachable call graph. It therefore distinguishes an absent entry point from an API that is present in a harness but realizes little of its downstream behavior. \emph{(4) Function-level coverage} describes internal functions through their runtime hits, uncovered branches, and the closest public API and call distance, thereby mapping an internal deficit back to a harness-controllable entry path. \emph{(5) Branch-level coverage} records the hit count of each branch and groups consecutive unhit branches into blocked regions. It therefore identifies contiguous coverage gaps and supplies the source region that the analyzer should inspect next.

Let $C_t^{\mathrm{API}}$ denote API coverage after iteration $t$, and let $\operatorname{stagnant}_3(t)$ denote three consecutive iterations without an API-coverage increase:
\begin{equation*}
\operatorname{stagnant}_3(t) \equiv
\bigwedge_{i=0}^{2}\left(C_{t-i}^{\mathrm{API}}=C_{t-i-1}^{\mathrm{API}}\right).
\end{equation*}
The next coverage-analysis phase is selected by
\begin{equation}
P_t =
\begin{cases}
P_2, & C_t^{\mathrm{API}} \geq 0.90 \;\lor\; \operatorname{stagnant}_3(t),\\
P_1, & \text{otherwise},
\end{cases}
\label{eq:coverage_phase_selection}
\end{equation}
where $P_1$ and $P_2$ denote the API-surface and deep-state exploration phases, respectively. In either phase, the \textit{Coverage Analyzer} selects one critical coverage gap and emits a coverage \emph{blueprint}, rather than harness code. The blueprint records the selected target, its coverage evidence, and the intended exploration effect. The active phase determines whether this blueprint describes an under-explored API surface or a state transition that resolves an internal blocker.

\emph{API-surface Exploration.}
The analyzer ranks each public API $a$ lexicographically in descending order by $\langle I(a), U(a)\rangle$, where $I(a)$ is its harness-invocation count and $U(a)$ is the accumulated number of uncovered branches over its statically reachable call graph. It selects the highest-ranked APIs as anchors, then consults module- and file-level coverage to find additional APIs $o$ that retain large $U(o)$ values \agentaction{by LLMs} and share a semantic or logical relationship with each anchor, such as operating on the same object or implementing adjacent operations in one lifecycle. The analyzer plans each resulting cluster as a single exploration unit and outputs a surface blueprint \agentaction{by LLMs} containing the target API list, their relationships, and the coverage potential motivating their joint exploration. This organization allows one subsequent harness to expand a coherent API surface instead of spending separate generation attempts on unrelated entry points.

\emph{Deep-state Exploration.}
The analyzer selects the largest contiguous branch blocker and invokes call-chain analysis to enumerate reverse paths from the containing internal function to public APIs, ordered from the shortest to the longest. For each candidate path, it reasons backward from the blocker using coverage-annotated source code with branch hit counts, identifying the predicate \agentaction{by LLMs} that stops execution and the API-induced state transitions capable of changing it~\cite{fuzzing_2023_fuzz_blockers}. If the shortest path cannot control the blocking condition, the analyzer proceeds to the next path. In this process, \sysname limits its intervention to blocking conditions that can be resolved through API invocations. The resulting blueprint records the diagnosed root cause and the ordered API sequence needed to remove the API-level blocker and establish a reachable path from the fuzzer entry to the target region. 
In the \tool{zlib} example from \autoref{fig:motivating_harness}, this process moves beyond the already covered \texttt{deflate} API, attributes the unhit region to the fixed \texttt{s->gzhead == Z\_NULL} state, and derives \texttt{deflateInit2} $\rightarrow$ \texttt{deflateSetHeader} $\rightarrow$ \texttt{deflate} as the sequence that can enter the blocked region.

\subsubsection{Dependency-Aware Harness Generation}
A coverage blueprint specifies the behavior to explore, but not an executable means of reaching it. The \textit{Harness Generator} realizes the blueprint by grounding it in declarations, implementations, data types, documentation, call sites, and available consumer code. Rather than treating the target APIs as independent calls, it models them as operations over library-managed state and resolves four kinds of dependency: construction dependencies that create valid objects, data-flow dependencies that propagate objects, temporal and state dependencies that constrain invocation order, and ownership dependencies that govern resource lifetimes. These relations turn the blueprint into a dependency-complete \emph{setup $\rightarrow$ target operations $\rightarrow$ teardown} lifecycle containing the helper calls and error paths needed to make the target behavior reachable.

Dependency recovery is guided jointly by the blueprint's target APIs and expected behavior. The API list specifies which operations must be realized, whereas the behavioral intent identifies the state transition or internal effect that their composition should induce. The generator uses this intent to focus source inspection and online knowledge retrieval to understand the relevant contracts \agentaction{by LLMs}. \emph{(1) Construction dependencies} are identified by tracing required objects backward to their allocators, factories, or initializers, together with the configuration needed to create a valid object. \emph{(2) Data-flow dependencies} connect return values and output parameters to downstream arguments. \emph{(3) Temporal and state dependencies} are recovered from implementation guards, documented ordering constraints, and recurring call sequences in consumers, revealing which transitions must precede or follow the target. \emph{(4) Ownership dependencies} pair acquisition, transfer, and release operations and follow normal and error-handling paths to determine who owns each resource and how long it must remain live. Combining implementation constraints with concrete consumer usage avoids mistaking type compatibility alone for a valid API composition~\cite{fuzzgen,graphfuzz}.

We represent these relations as a heterogeneous graph $G=(V_A, E_C \cup E_D \cup E_S \cup E_O)$, where $V_A$ contains API instances and the four edges encode construction, data flow, state ordering, and ownership. Starting from the blueprint's target nodes, the generator computes a dependency closure by expanding backward until every resource requirement and API constraint is satisfied \agentaction{by LLMs}. Calls that do not contribute to this closure are discarded. Finally, the generator linearizes the graph into harness code \agentaction{by LLMs}, by instantiating API instances and constructing the setup, teardown, and error-handling paths needed to satisfy the dependencies. The resulting harness makes the intended transition reachable at the API level to efficiently explore API surfaces or deep branches.

\subsection{Evidence-Based Crash Validation}
\label{sec:crash_driven_evolution}

\sysname treats crash validation as a process of constructing evidence rather than classifying a sanitizer report by plausibility in appearance. A call trace identifies where a failure manifests, but it does not establish whether the library received a valid object or which usage rules applied to that execution. \sysname therefore answers two questions before issuing a verdict: what concrete state leads to the crash, and whether that state satisfies the applicable contract. This yields the diagnostic pipeline motivated in \autoref{sec:background:crash_validation_example}: \emph{reproduce $\rightarrow$ recover runtime state $\rightarrow$ reconstruct the applicable contract $\rightarrow$ derive an evidence-grounded verdict}.

\noindent
\textbf{Runtime-state recovery.}
The \textit{Crash Analyzer} first invokes the Crash Debugging Interface to minimize the harness while preserving the failure signature and to reproduce the failure with its crashing input. The sanitizer trace localizes the manifestation point; the analyzer then selects the runtime values needed to distinguish a valid library state from a harness misuse at the relevant frames \agentaction{by LLMs}. The interface extracts their concrete values together with the associated object fields, buffer boundaries, and source locations at one execution state. These observations form structured snapshots of the object states, connecting the runtime values to both the failing locations and their harness-controlled origin. In the \tool{libaom} example, this snapshot exposes the selected value of \texttt{img.fmt}, the corresponding plane metadata and the buffer offsets at the overflow point that remain ambiguous in the harness shown in \autoref{fig:motivating_crash_harness}.

\noindent
\textbf{Adaptive contract reconstruction.}
The recovered snapshots make context construction state-directed rather than bounded by the failing trace. The analyzer uses the observed discriminants and their dependent fields to trace how the object was constructed and consumed \agentaction{by LLMs}. From this state, it retrieves the constraints governing the concrete execution, including those defined outside the functions on the trace. It then checks each constraint against the recovered values and classifies it as satisfied, violated, or unresolved \agentaction{by LLMs}. Unresolved constraints guide the next diagnostic step by identifying which runtime value or source context is still missing. The analyzer continues gathering this evidence until it can support a classification instead of forcing a one-step verdict. When the evidence ties a contract violation to a harness operation, the analyzer reports harness misuse and returns an actionable correction. When the library mishandles a contract-valid state, it retains the failure as a candidate library bug. Adaptive reconstruction therefore provides more evidence than static context and grounds each verdict in rules.


\begin{table*}[htp]
    \centering
    \fontsize{6.6pt}{7.4pt}\selectfont
    \sisetup{detect-weight=true}
    \setlength{\tabcolsep}{1pt} 
    \newcommand{\libhead}[1]{{\fontsize{6.0pt}{6.8pt}\selectfont\textbf{#1}}}
    
    \begin{threeparttable}
        \caption{Overview of the evaluation results across 20 libraries.}
        \label{tab:eval_overview}
        
        \begin{tabular}{l *{20}{c} c}
            \toprule
             & \libhead{cJSON} & \libhead{libmagic} & \libhead{RE2} & \libhead{pugixml} & \libhead{zlib} & \libhead{c-ares} & \libhead{liblouis} & \libhead{libpng} & \libhead{libpcap} & \libhead{libtiff} & \libhead{lcms} & \libhead{tinygltf} & \makecell{\libhead{libjpeg}\\\libhead{-turbo}} & \libhead{curl} & \libhead{libvpx} & \libhead{SQLite3} & \libhead{libaom} & \libhead{protobuf} & \libhead{OpenSSL} & \libhead{OpenCV} & \libhead{Total} \\
            
            \midrule
            \multicolumn{22}{l}{\textit{\textbf{Library Statistics}}} \\
            \midrule
            Language     & C       & C        & C++     & C++     & C       & C       & C        & C       & C       & C       & C       & C++      & C             & C        & C       & C       & C       & C++      & C       & C++     &  -        \\
            Version      & b2890   & 772f2    & 972a1   & 71005   & 09a15   & 2870f   & e90a9    & 7c67f   & a516c   & 5fe20   & 36039   & bdc37    & 466c3         & e8415a   & cb5a5   & db4d8   & 16a97   & b56a4    & c8b4a   & 01f9b   &  -        \\
            LoC          & 13K     & 19K      & 33K     & 42K     & 44K     & 52K     & 65K      & 108K    & 122K    & 150K    & 154K    & 194K     & 212K          & 314K     & 501K    & 580K    & 857K    & 1.26M    & 1.58M   & 3.19M   & 9.49M        \\
            \#APIs       & \num{78}      & \num{18}       & \num{108}     & \num{352}     & \num{88}      & \num{138}     & \num{34}       & \num{271}     & \num{108}     & \num{191}     & \num{296}     & \num{466}      & \num{146}           & \num{156}      & \num{37}      & \num{298}     & \num{47}      & \num{3149}     & \num{6728}    & \num{4236}    & \num{16945}        \\
            \#Branches   & \num{1060}    & \num{7802}     & \num{5060}    & \num{4448}    & \num{3038}    & \num{9172}    & \num{7338}     & \num{7810}    & \num{7394}    & \num{15466}   & \num{9180}    & \num{7584}     & \num{11388}         & \num{19980}    & \num{33082}   & \num{54600}   & \num{63124}   & \num{25498}    & \num{130338}  & \num{237392}  & \num{660754}         \\
            \midrule
            \multicolumn{22}{l}{\textit{\textbf{Metric 1: Branch Coverage}}} \\
            \midrule
            OSS-Fuzz     & \num{489}     & \num{3833}     & \num{2602}    & \num{2489}    & \num{1959}    & \num{3340}    & \num{4687}     & \num{1575}    & \num{3106}    & \num{3078}    & \num{4110}    & \num{1101}     & \num{7363}          & \num{5126}     & \num{11677}   & \num{17690}   & \num{11122}   & \num{477}      & \textbf{\num{35983}}   & \num{1985}    & \textbf{\num{123792}}   \\
            OSS-Fuzz-Gen & \num{694}     & \num{4103}     & \num{2396}    & \num{2786}    & \num{2040}    & \num{2025}    & \num{1778}     & \num{1378}    & \num{2428}    & \num{3673}    & \num{1976}    & \num{157}      & \num{1944}          & \num{892}      & \num{2915}    & \num{15458}   & \num{9459}    & \num{696}      & \num{1805}    & \num{3075}    & \textbf{\num{61678}}    \\
            PromptFuzz   & \num{831}     & \num{4639}     & \num{1780}    & -       & \num{2368}    & \num{5326}    & \num{4428}     & \num{3361}    & \num{3804}    & \num{7049}    & \num{4138}    & -        & \num{3360}          & \num{5699}     & \num{9290}    & \num{18454}   & \num{18382}   & -        & \num{10777}   & -       & \textbf{\num{103686}}   \\
            PromeFuzz    & \num{805}     & \num{3905}     & \num{3253}    & \num{3316}    & \num{1985}    & \num{5091}    & \num{3624}     & \num{1849}    & \num{4239}    & \num{7586}    & \num{3673}    & \num{1925}     & \num{4975}          & \num{5596}     & \num{6558}    & \num{13041}   & \num{22086}   & -        & -       & -       & \textbf{\num{93507}}    \\
                        ProteusFuzz  & \num{871}     & \num{3970}     & \num{2597}     & \num{3307}     & \num{2186}     & \num{1435}     & \num{1599}     & \num{3506}     & \num{3316}     & \num{7599}     & \num{3934}     & \num{3178}     & \num{4876}     & \textbf{\num{9263}}     & \num{6454}     & \num{17077}     & \num{20533}     & -     & -     & -     & \textbf{\num{95702}}    \\
            FuzzAgent (E2E)    & \num{880}     & \num{4484}     & \num{4069}    & \num{3582}    & \num{2598}    & \num{5182}    & \num{4911}     & \num{4921}    & \num{5120}    & \num{8212}    & \num{4537}    & \num{3056}     & \num{7541}          & \num{6699}     & \num{15680}   & \num{26197}   & \num{33308}   & \num{7664}     & \num{18227}   & \num{12751}   & \textbf{\num{179619}}   \\
            FuzzAgent   & \textbf{\num{882}}     & \textbf{\num{4927}}     & \textbf{\num{4174}}    & \textbf{\num{3757}}    & \textbf{\num{2613}}    & \textbf{\num{5387}}    & \textbf{\num{5078}}     & \textbf{\num{5342}}    & \textbf{\num{5455}}    & \textbf{\num{9139}}    & \textbf{\num{4681}}    & \textbf{\num{3631}}     & \textbf{\num{7713}}          & \num{6117}     & \textbf{\num{16400}}   & \textbf{\num{26160}}   & \textbf{\num{33832}}   & \textbf{\num{7789}}   & \num{18374}   & \textbf{\num{13389}}   & \textbf{\num{184840}} \\
            \midrule
            \multicolumn{22}{l}{\textit{\textbf{Metric 2: LLM Cost}}} \\
            \midrule
            OSS-Fuzz-Gen & \$21.62 & \$11.90  & \$11.37 & \$3.59  & \$3.44  & \$14.22 & \$11.51  & \$13.71 & \$10.45 & \$12.53 & \$14.98 & \$3.56   & \$6.42        & \$2.88   & \$13.94 & \$12.85 & \$9.17  & \$22.49  & \$1.08  & \$1.73  & \textbf{\$203.44} \\
            PromptFuzz   & \$0.41  & \$0.86   & \$2.78  & -       & \$0.44  & \$2.63  & \$1.46   & \$2.84  & \$0.54  & \$2.68  & \$1.45  & -        & \$1.18        & \$1.90   & \$0.73  & \$1.24  & \$0.35  & -        & \$1.66  & -       & \textbf{\$23.15}    \\
            PromeFuzz    & \$0.22  & \$0.05   & \$0.34  & \$0.62  & \$0.28  & \$1.08  & \$0.38   & \$2.66  & \$0.34  & \$1.13  & \$2.11  & \$0.22   & \$0.37        & \$0.46   & \$0.33  & \$5.39  & \$0.68  & \$223.06 & \$221.80 & \$214.49  &  \textbf{\$676.01} \\
            ProteusFuzz  & \$0.24  & \$0.05   & \$0.75  & \$0.45  & \$0.54  & \$1.76  & \$0.52   & \$2.19  & \$0.95  & \$0.54  & \$1.02  & \$4.03   & \$0.34        & \$2.31   & \$1.57  & \$1.52  & \$1.88  & \$223.06 & \$221.80 & \$214.49  &  \textbf{\$680.01} \\
            FuzzAgent    & \$3.51  & \$2.65   & \$2.34  & \$2.50  & \$3.13  & \$3.31  & \$3.27   & \$3.83  & \$2.66  & \$3.54  & \$3.51  & \$3.60   & \$3.57        & \$3.31   & \$3.36  & \$3.12  & \$2.98  & \$3.73   & \$2.77  & \$2.75  & \textbf{\$63.44}    \\

            \midrule
            \multicolumn{22}{l}{\textit{\textbf{Metric 3: Detected Bugs}}} \\
            \midrule
            OSS-Fuzz     & \num{0}       & \num{0}        & \num{0}       & \num{0}       & \num{0}       & \num{0}       & \num{0}        & \num{0}       & \num{0}       & \num{0}       & \num{0}       & \num{0}        & \num{0}             & \num{0}        & \num{0}       & \num{0}       & \num{0}       & \num{0}        & \num{0}       & \num{0}       &   \num{0}       \\
            OSS-Fuzz-Gen        & \num{0}       & \num{0}        & \num{0}       & \num{0}       & \num{0}       & \num{0}       & \num{0}        & \num{0}       & \num{0}       & \num{0}       & \num{0}       & \num{0}        & \num{0}             & \num{0}        & \num{0}       & \num{0}       & \num{0}       & \num{0}        & \num{0}       & \num{0}       &   \num{0}       \\
            PromptFuzz   & \num{1}       & \num{0}        & \num{0}       & -       & \num{0}       & \num{0}       & \num{1}        & \num{0}       & \num{0}       & \num{0}       & \num{0}       & -        & \num{1}             & \num{0}        & \num{2}       & \num{0}       & \num{0}       & -        & \num{0}        & -        & \num{5}        \\
            PromeFuzz    & \num{0}       & \num{0}        & \num{0}       & \num{0}       & \num{0}       & \num{0}       & \num{0}        & \num{0}       & \num{1}       & \num{0}       & \num{0}       & \num{0}        & \num{0}             & \num{0}        & \num{0}       & \num{0}       & \num{0}       & -        & -        & -        & \num{1}        \\
            ProteusFuzz  & \num{0}       & \num{0}        & \num{0}       & \num{0}       & \num{0}       & \num{0}       & \num{0}        & \num{0}       & \num{0}       & \num{1}       & \num{1}       & \num{0}        & \num{0}             & \num{0}        & \num{1}       & \num{0}       & \num{1}       & -        & -        & -        & \num{4}        \\
            FuzzAgent    & \num{3}     & \num{2}      & \num{0}       & \num{3}     & \num{1}     & \num{1}     & \num{8}      & \num{2}     & \num{2}     & \num{1}     & \num{3}     & \num{2}        & \num{3}           & \num{0}        & \num{28}   & \num{1}     & \num{21}   & \num{2}      & \num{6}     & \num{13}     & \textbf{\num{102}}   \\
            \bottomrule
        \end{tabular}

        \begin{tablenotes}[flushleft]
            \item \textbf{LoC}: Lines of Code. \textbf{FuzzAgent (E2E)}: evaluation results of \sysname within only 24-hour end-to-end execution, without 24 hours of standalone fuzzing. \textbf{PromptFuzz} does not support C++ (\texttt{RE2} is evaluated through its \texttt{cre2} C wrapper). \textbf{ProteusFuzz} reuses the preprocessing of \tool{PromeFuzz}, so neither could complete it within 24 hours for \texttt{protobuf}, \texttt{OpenSSL}, and \texttt{OpenCV}. Unsupported or unevaluated entries are marked ``-''.
        \end{tablenotes}
        \end{threeparttable}
\end{table*}

\section{Implementation}
\label{sec:impl}

\sysname is implemented in 17k lines of Python code. Its agent runtime coordinates a persistent workspace and exposes typed tools following the modularity and robustness principles recommended for agentic systems~\cite{openai_practical_guide_building_agents,modelcontext_writing_effective_tools}. File navigation and editing tools are adapted from \tool{SWE-Agent}~\cite{yang2024sweagent}; the Web Interface uses the GitHub API~\cite{github_api}; and the Fuzzing Interface integrates build, compilation, and fuzzing scripts implemented specifically for library fuzzing. The Coverage Analysis Interface uses source-based code coverage~\cite{source_based_code_coverage} and \tool{llvm-cov}~\cite{llvm-cov-tool} to extract the coverage hierarchy, and implements call graph analysis by tree-sitter~\cite{tree_sitter} in scripts.
For the Crash Debugging Interface, it uses \tool{CASR}~\cite{savidov2021casr} for crash reproduction, \tool{tree-reduce}~\cite{tree_reduce} for harness minimization and custom scripts with \tool{GDB}~\cite{gdb_batch_model} for runtime state inspection. 
The stateful environment is implemented by agent hooking~\cite{claude_reminder}.

\section{Evaluation}
\label{sec:eval}

This section presents a comprehensive evaluation of \sysname across 20 widely-used C/C++ libraries. To rigorously assess \sysname's efficiency and effectiveness, we compare it against five representative baselines: \tool{OSS-Fuzz}~\cite{oss_fuzz}, \tool{OSS-Fuzz-Gen}~\cite{oss-fuzz-gen}, \tool{PromptFuzz}~\cite{ccs_24_promptfuzz}, \tool{PromeFuzz}~\cite{ccs_25_promefuzz} and \tool{ProteusFuzz}~\cite{tdsc_26_proteusfuzz}, covering the spectrum from an industrial continuous fuzzing infrastructure to state-of-the-art LLM-driven harness generators. The evaluation targets two complementary groups of libraries: 17 representative libraries previously evaluated by \tool{PromptFuzz},  \tool{PromeFuzz} and \tool{ProteusFuzz}, enabling direct and fair comparison, and 3 large-scale, industrially critical libraries (\tool{protobuf}, \tool{OpenSSL}, and \tool{OpenCV}) whose complex codebases pose challenges to real-world deployment.

All experiments were conducted on a server equipped with two Intel Xeon Platinum 8580 processors (120 physical cores / 240 threads, 2.0\,GHz), 8 NVIDIA H200 GPUs, and 2TB of RAM, running Ubuntu 24.04 LTS. All systems were evaluated under \tool{DeepSeek V3.2}~\cite{deepseek_v32} and allocated 24 hours for their executions and 24 hours for standalone fuzzing. The system-specific settings are provided in \apref{sec:system-specific-setting}. To reduce the statistical variance in library fuzzing as observed in \autoref{sec:discuss}, we repeated five independent execution rounds for each system. With each round, we repeated the subsequent 24-hour standalone fuzzing campaign five times.

\subsection{Automation}
\label{sec:automation-efficiency}

We evaluate automation as an end-to-end system property rather than as harness generation alone. We examine two questions: whether the lifecycle completes without human intervention and whether the automatically acquired domain inputs remain useful. We additionally evaluated the dominance of domain-specific operations and internal time utilization in \sysname, and provide the results in \apref{sec:automation-details}.

\subsubsection{End-to-End Completion and Human Effort}

We measure human onboarding from the same source-sufficient starting point. Four PhD students performed and logged the baseline setup. The measured person-hours start from preparing the fuzzing setup with only source code to launch a valid fuzzing campaign. In contrast, \sysname receives the source tree without any of these prepared artifacts.

\begin{table*}[t]
    \centering
    \fontsize{6.2pt}{7.0pt}\selectfont
    \setlength{\tabcolsep}{0.9pt}
    \sisetup{detect-weight=true}
    \newcommand{\libhead}[1]{{\fontsize{5.8pt}{6.4pt}\selectfont\textbf{#1}}}
    \begin{threeparttable}
        \caption{Utility of dictionary and seed generation across \num{20} libraries.}
        \label{tab:seed_dict_ablation}
        \begin{tabular}{l *{20}{r} rr}
            \toprule
            \textbf{Configuration} & \libhead{cJSON} & \libhead{libmagic} & \libhead{RE2} & \libhead{pugixml} & \libhead{zlib} & \libhead{c-ares} & \libhead{liblouis} & \libhead{libpng} & \libhead{libpcap} & \libhead{libtiff} & \libhead{lcms} & \libhead{tinygltf} & \makecell{\libhead{libjpeg}\\\libhead{-turbo}} & \libhead{curl} & \libhead{libvpx} & \libhead{SQLite3} & \libhead{libaom} & \libhead{protobuf} & \libhead{OpenSSL} & \libhead{OpenCV} & \libhead{Total} & \libhead{Gain} \\
            \midrule
            Empty            & \num{858} & \num{3729} & \num{4926} & \num{3097} & \num{2232} & \num{4733} & \num{939} & \num{2072} & \num{4098} & \num{2817} & \num{2135} & \num{2337} & \num{3883} & \num{4053} & \num{8661} & \num{14760} & \num{18816} & \num{6448} & \num{15569} & \num{11109} & \num{117273} & -- \\
            OSS dictionary   & \num{865} & \num{3682} & \num{4915} & \num{3080} & \num{2098} & \num{4924} & \num{967} & \num{2066} & \num{4200} & \num{2494} & \num{2350} & \num{2331} & \num{3876} & \num{3949} & \num{8655} & \num{14634} & \num{17888} & \num{6388} & \num{15667} & \num{10059} & \num{115087} & \num{-1.86}\% \\
            \textbf{Agent dictionary} & \num{866} & \num{3812} & \num{4998} & \textbf{\num{3523}} & \num{2192} & \num{4931} & \num{1213} & \num{2234} & \textbf{\num{5082}} & \num{3739} & \num{2564} & \num{2385} & \textbf{\num{6058}} & \num{5579} & \num{7781} & \textbf{\num{20016}} & \num{20305} & \num{6615} & \num{15899} & \textbf{\num{11286}} & \num{131077} & \textbf{\num{11.77}\%} \\
            OSS seeds        & \num{868} & \num{3283} & \num{4982} & \num{3342} & \num{2302} & \textbf{\num{5194}} & \num{974} & \num{4396} & \num{4799} & \num{5631} & \textbf{\num{2926}} & \num{2331} & \num{4383} & \textbf{\num{5582}} & \num{10218} & \num{18795} & \textbf{\num{24874}} & \num{6598} & \num{15829} & \num{9802} & \num{137110} & \num{16.91}\% \\
            \textbf{Agent seeds}      & \textbf{\num{869}} & \textbf{\num{4029}} & \textbf{\num{5024}} & \num{3398} & \textbf{\num{2305}} & \num{4921} & \textbf{\num{1578}} & \textbf{\num{4648}} & \num{4584} & \textbf{\num{7148}} & \num{2752} & \textbf{\num{2405}} & \num{5175} & \num{5395} & \textbf{\num{10656}} & \num{19180} & \num{22194} & \textbf{\num{7010}} & \textbf{\num{16295}} & \num{11034} & \num{140597} & \textbf{\num{19.88}\%} \\
            \bottomrule
        \end{tabular}
        \begin{tablenotes}[flushleft]
            \item \textbf{Note:} Each cell reports mean branch coverage after five hours over five runs on \sysname's fuzzers. \textit{Empty} uses no dictionary and no seed corpus.
        \end{tablenotes}
    \end{threeparttable}
\end{table*}

\begin{table}[t]
    \centering
    \footnotesize
    \setlength{\tabcolsep}{2.5pt}
    \begin{threeparttable}
        \caption{Source-sufficient onboarding effort for each system.}
        \label{tab:automation_boundary}
        \begin{tabular*}{\columnwidth}{@{\extracolsep{\fill}}lccccr@{}}
            \toprule
            \textbf{System} & \textbf{Build} & \textbf{Dict \& seed} & \textbf{Config} & \textbf{Success} & \textbf{Hours} \\
            \midrule
            \tool{OSS-Fuzz-Gen} & M & M & M & 20/20 & $\sim$ 5 \\
            \tool{PromptFuzz}   & M & M & M & 16/20 & $\sim$ 8 \\
            \tool{PromeFuzz}    & M & M & M & 17/20 & $\sim$ 12 \\
            \tool{ProteusFuzz}    & M & M & M & 17/20 & $\sim$ 12 \\
            \sysname            & A & A & A & \textbf{20/20} & \textbf{0} \\
            \bottomrule
        \end{tabular*}
        \begin{tablenotes}[flushleft]
          \item  \textbf{Build}: build and instrumentation; \textbf{Dict \& seed}: dictionaries and seeds; \textbf{Config}: system-specific configuration; \textbf{A}: automated; \textbf{M}: manual.
        \end{tablenotes}
    \end{threeparttable}
\end{table}

As shown in \autoref{tab:automation_boundary}, \sysname completed the full lifecycle for all 20 libraries in each of the five end-to-end rounds, yielding 100/100 completed library--rounds without human intervention. The generation-based baselines automate harness synthesis only after their operating environment has been prepared manually. Even for targets included in their original evaluations, reproducing this preparation required approximately one person-hour per library. For newly added targets, the average increased to 5 hours for \tool{OSS-Fuzz-Gen}, 8 hours for \tool{PromptFuzz}, and 12 hours each for \tool{PromeFuzz} and \tool{ProteusFuzz}. The work involved complex build systems, dependency conflicts, auxiliary inputs, and system-specific materials such as consumer code and documentation; the dependence of \tool{PromptFuzz} on such manual preparation is also documented by its developers~\cite{promptfuzz_author_response}.

Manual configuration did not guarantee that the automated portion would complete. In particular, \tool{PromeFuzz}'s codebase summarization, which \tool{ProteusFuzz} reuses, failed to finish within 24 hours on \tool{protobuf}, \tool{OpenSSL}, and \tool{OpenCV}, despite costing more than \$200 per target. This behavior is consistent with its $O(n^2)$ pairwise summary construction over $n$ API functions~\cite{promefuzz_issue}. In contrast, \sysname has no pairwise preprocessing stage and its inference cost scales only with the running time and averages \$3.17 per target for the complete end-to-end workflow. Thus, \sysname extends the automation boundary in two respects: it internalizes the project-specific setup normally assigned to an expert, and it sustains that workflow on large targets for which a baseline can stall before harness generation begins.

\subsubsection{Utility of Autonomously Acquired Inputs}

A source-sufficient system must recover the input-domain knowledge that human operators normally encode in dictionaries and seed corpora. \sysname acquires this knowledge from related projects and filters the retrieved artifacts against the target's formats and protocols. We evaluate their utility through five five-hour fuzzing runs under each input configuration. Each run was evaluated on the 100 fuzzers generated and merged by \sysname across five rounds on 20 libraries.

As shown in \autoref{tab:seed_dict_ablation}, fuzzing without either artifact covers \num{117273} branches. The dictionary acquired by \sysname increases this total by 11.77\% to \num{131077}, whereas the OSS-Fuzz dictionary does not improve aggregate coverage in this setting. The seed corpus acquired by \sysname reaches \num{140597} branches, a 19.88\% gain over the empty configuration. Relative to their OSS-Fuzz counterparts, the agent-acquired dictionary and seeds cover 13.89\% and 2.54\% more branches, respectively. The automatically prepared artifacts are therefore not merely present or syntactically valid; their measured utility matches or exceeds the manually curated resources supplied to the baselines.




\subsection{Code Coverage}
\label{sec:coverage_effectiveness}

We evaluate coverage effectiveness from three complementary perspectives. We first compare the aggregate branch coverage achieved by each system under the common experimental protocol. We then examine how efficiently generated harnesses survive validation and translate into API, function, and branch coverage, capturing both exploration breadth and depth. Finally, we ablate dependency-aware harness generation and the \textit{Coverage Analyzer} to quantify the contribution of each component.

\subsubsection{Overall Coverage Effectiveness}

For a direct comparison, we use the 15 libraries completed by all six systems. Under the common evaluation protocol, \sysname covers \num{137900} branches, compared with \num{92909} for \tool{PromptFuzz}, \num{89216} for \tool{ProteusFuzz}, \num{88266} for \tool{PromeFuzz}, \num{81757} for \tool{OSS-Fuzz}, and \num{53159} for \tool{OSS-Fuzz-Gen}. These totals represent improvements of 48.4\%, 54.6\%, 56.2\%, 68.7\%, and 159.4\%, respectively. \sysname achieves the highest coverage on 14 of the 15 shared targets; the exception is \tool{curl}, where \tool{ProteusFuzz} covers \num{9263} branches against \num{6117}.
Without the additional standalone fuzzing phase, \textit{FuzzAgent (E2E)} already covers \num{134339} branches on the same 15 libraries, only 2.6\% below its result after standalone fuzzing. Across the complete 20-library benchmark, it covers \num{179619} branches, exceeding \tool{OSS-Fuzz}, \tool{PromptFuzz}, \tool{ProteusFuzz}, \tool{PromeFuzz}, and \tool{OSS-Fuzz-Gen} by 45.1\%, 73.2\%, 87.7\%, 92.1\%, and 191.2\%, respectively. The improvements are statistically significant for most library--baseline pairs under the Mann--Whitney U tests reported in \apref{sec:mann-whitney-u-test-for-statistical-significance}. Thus, \sysname achieves consistently high coverage across the benchmark and obtains nearly all of its final coverage within the end-to-end evolutionary workflow.

\subsubsection{Harness Efficiency and Exploration Depth}
\label{sec:harness-efficiency}
\begin{table}[t]
    \centering
    \setlength{\tabcolsep}{1.5pt}
    \sisetup{detect-weight=true}
    \begin{threeparttable}
        \caption{Harness-to-branch-coverage conversion efficiency across five end-to-end runs per evaluated library.}
        \label{tab:harness_conversion}
        \footnotesize
        \begin{tabular}{lr@{\hspace{4pt}}r@{\hspace{4pt}}rrr}
            \toprule
            \makecell[c]{\textbf{System}} & \makecell[c]{\textbf{Gen.}} & \makecell[c]{\textbf{Retained}} & \makecell[c]{\textbf{Coverage}\\\textbf{Effective}} & \makecell[c]{\textbf{Gain per}\\\textbf{Gen.}} & \makecell[c]{\textbf{Gain per}\\\textbf{Effective}} \\
            \midrule
            \tool{PromptFuzz} & \num{55251} & \num{6933} (\num{12.55}\%) & \num{3675} (\num{6.65}\%) & \num{9.17} & \num{137.90} \\
            \tool{PromeFuzz}  & \num{17987} & \num{4958} (\num{27.56}\%) & \num{3771} (\num{20.97}\%) & \num{28.80} & \num{137.38} \\
            \tool{ProteusFuzz} & \num{31224} & \num{6439} (\num{20.62}\%) & \num{4530} (\num{14.51}\%) & \num{15.33} & \num{105.63} \\
            \sysname          & \num{2646}  & \num{1582} (\textbf{\num{59.79}\%}) & \num{1428} (\textbf{\num{53.97}\%}) & \textbf{\num{330.29}} & \textbf{\num{612.01}} \\
            \bottomrule
        \end{tabular}
        \begin{tablenotes}[flushleft]
            \item \textbf{Gen.}: Generation attempts, including repairs. \textbf{Retained}: harnesses passing each system's validation. \textbf{Coverage-Effective}: harnesses adding a previously uncovered target branch in retained order. Parentheses give percentages of attempts. \textbf{Gain per Gen./Effective}: Total E2E target branches divided by the respective counts.
        \end{tablenotes}
    \end{threeparttable}
\end{table}

\begin{figure*}[!t]
    \centering
    \includegraphics[width=1\linewidth]{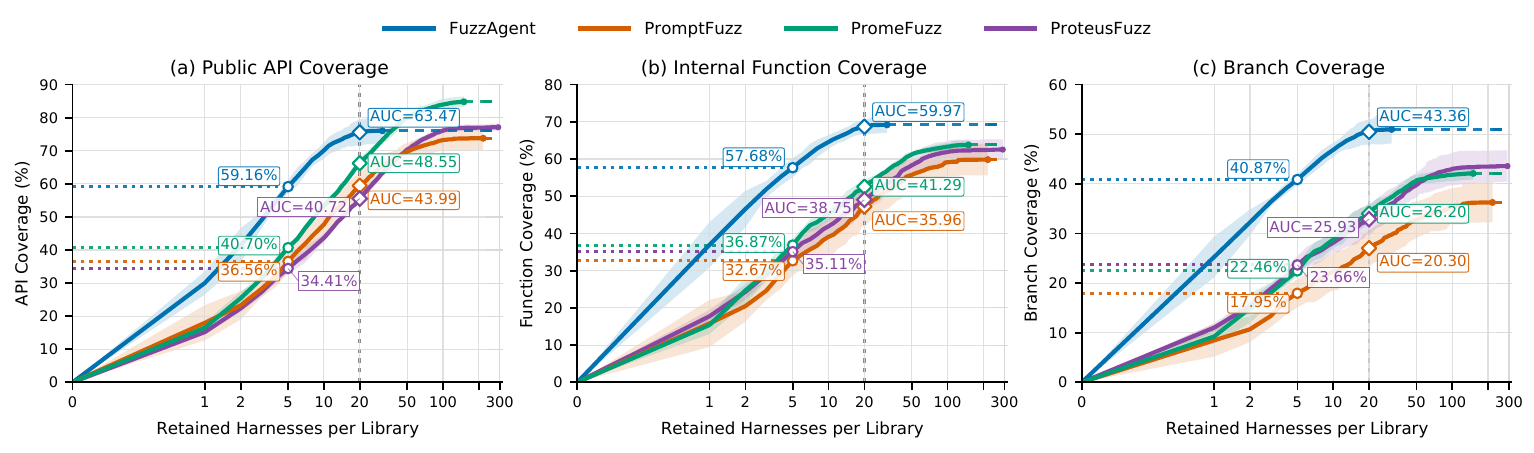}
    \caption{\small{Cumulative coverage over retained harnesses on the 15 libraries shared by \sysname, \tool{PromptFuzz}, \tool{PromeFuzz}, and \tool{ProteusFuzz}. Solid lines average five E2E rounds, each macro-averaged across libraries; shaded bands show 95\% t confidence intervals. Solid curves terminate at the maximum retained-harness budgets (31, 218, 150, and 291, respectively). Dashed tails are visual guides.}}
    \label{fig:harness_coverage_efficiency}
\end{figure*}

To characterize how this effectiveness emerges at the harness level, we trace each system's generated candidates through validation and coverage contribution in \autoref{tab:harness_conversion}.
Only 12.55\% of \tool{PromptFuzz}'s generated harnesses survive validation, and 6.65\% contribute unique branch coverage. The corresponding rates improve to 27.56\% and 20.97\% for \tool{PromeFuzz} and 20.62\% and 14.51\% for \tool{ProteusFuzz}, but reach 59.79\% and 53.97\% for \sysname. The difference persists after accounting for generation effort: each \sysname attempt adds 330.29 target branches on average, compared with 9.17 for \tool{PromptFuzz}, 28.80 for \tool{PromeFuzz}, and 15.33 for \tool{ProteusFuzz}. Even among coverage-effective harnesses, \sysname gains 612.01 branches per harness on average, versus approximately 138 for \tool{PromptFuzz} and \tool{PromeFuzz} and 105.63 for \tool{ProteusFuzz}. The results therefore show that \sysname's coverage advantage comes from converting larger generation attempts into valid, coverage-contributing harnesses, rather than from generating more candidates. \tool{ProteusFuzz} retains \num{6439} of \num{31224} attempts, which include repairs and adaptive candidates. It makes \num{4530} of them coverage-effective, but each adds the fewest branches of the four systems.

\autoref{fig:harness_coverage_efficiency} traces cumulative coverage over the complete sequence of harnesses that survive each system's validation pipeline, thereby capturing the coverage productivity of its retained harness population. At a budget of five retained harnesses per library round, \sysname already covers 59.16\% of APIs, 57.68\% of internal functions, and 40.87\% of branches. The corresponding results are 36.56\%, 32.67\%, and 17.95\% for \tool{PromptFuzz}, 40.70\%, 36.87\%, and 22.46\% for \tool{PromeFuzz}, and 34.41\%, 35.11\%, and 23.66\% for \tool{ProteusFuzz}. Over the first 20 harnesses, \sysname likewise yields the largest normalized AUC at all three levels (63.47, 59.97, and 43.36 percentage points), compared with 43.99, 35.96, and 20.30 for \tool{PromptFuzz}, 48.55, 41.29, and 26.20 for \tool{PromeFuzz}, and 40.72, 38.75, and 25.93 for \tool{ProteusFuzz}. This advantage emerges early because \sysname organizes semantically related target APIs and their dependencies into coherent harnesses, allowing each retained harness to exercise multiple entry points and unlock their downstream states instead of exploring APIs through isolated generation attempts.

At the endpoints of \autoref{fig:harness_coverage_efficiency}, \sysname reaches 76.03\% API, 69.19\% function, and 50.97\% branch coverage with a maximum budget of 31 retained harnesses per library--round. \tool{PromptFuzz}'s curve extends to 218 harnesses---7.03$\times$ this budget---but finishes 2.24, 9.34, and 14.71 percentage points lower at the respective levels. \tool{PromeFuzz}'s curve extends to 150 harnesses, 4.84$\times$ the budget of \sysname. At this larger budget, it reaches the highest API coverage (84.82\%, 8.79 points above \sysname), but only 63.80\% function and 42.10\% branch coverage, 5.39 and 8.87 points below \sysname. \tool{PromptFuzz} therefore fails to match \sysname at any level despite its much larger budget, whereas \tool{PromeFuzz} converts scale primarily into API-surface breadth. \tool{ProteusFuzz}'s curve extends to 291 harnesses, 9.39$\times$ the budget of \sysname, and reaches 77.11\% API coverage, 1.08 points above \sysname. Its function and branch coverage are 62.55\% and 43.58\%, 6.64 and 7.39 points below \sysname, so it shows the same breadth--depth gap despite a much larger harness population.


Taken together, \autoref{tab:harness_conversion} and \autoref{fig:harness_coverage_efficiency} show that \sysname's advantage stems from the diagnostic resolution of its feedback rather than from iterative generation alone. \tool{PromptFuzz} and \tool{PromeFuzz} primarily expose sparse, outcome-level signals: they indicate that a harness fails or that an API remains uncovered, but not why exploration is blocked or how the harness should change. Their generation consequently remains a broad generate-and-filter search over API combinations. \tool{ProteusFuzz} adds adaptive repair and blocker-driven harness synthesis, which raises its retention and coverage-effective rates above those of \tool{PromptFuzz}. Yet it still needs 291 retained harnesses to reach 43.58\% branch coverage, and its gain per effective harness is the lowest of the four systems. \sysname instead maintains a stateful campaign context that combines hierarchical coverage with library source and documentation, API dependencies, and prior execution outcomes. Its specialized agents diagnose concrete coverage blockers, translate them into invocation sequences with the necessary helper calls and constraints, adapt between API-surface expansion and deep-state exploration, and terminate low-yield campaigns at coverage plateaus. Each subsequent harness therefore tests a concrete hypothesis about the current bottleneck, explaining both the higher conversion rates and the steeper coverage curves.

\subsubsection{Ablation on Dependency-Aware Harness Generation}
\begin{table}[t]
    \centering
    \setlength{\tabcolsep}{4pt}
    \sisetup{detect-weight=true}
    \begin{threeparttable}
        \caption{Ablation of dependency-aware harness generation, averaged over \num{20} libraries.}
        \label{tab:dep_aware_ablation}
        \begin{tabular}{l rr rr}
            \toprule
            \multirow{2}{*}{\textbf{Configuration}} & \multicolumn{2}{c}{\textbf{Branch}} & \multicolumn{2}{c}{\textbf{Function}} \\
            \cmidrule(lr){2-3} \cmidrule(l){4-5}
            & \textbf{Covered} & \textbf{Gain} & \textbf{Covered} & \textbf{Gain} \\
            \midrule
            w/o dependency-aware         & 2247.6 & -- & 476.4 & -- \\
            \textbf{w/ dependency-aware} & 4584.6 & \textbf{104.0\%} & 860.8 & \textbf{80.7\%} \\
            \bottomrule
        \end{tabular}
    \end{threeparttable}
\end{table}

The harness-level comparison above does not isolate dependency-aware generation from the rest of the workflow, since the two also differ in feedback and scheduling. We therefore ablate this policy alone. For each library, we generate 10 harnesses with the policy enabled and 10 with it disabled. In each generation round, we randomly sample three target APIs and query the LLM until the harness compiles. Every compiled harness is then fuzzed for five hours, and we measure the branches and functions it covers. Both configurations use the same sampling procedure, compilation criterion, and fuzzing budget.

\autoref{tab:dep_aware_ablation} shows that disabling the policy roughly halves the coverage of a generated harness. Averaged over the 20 libraries, a harness covers 2247.6 branches and 476.4 functions without it, versus 4584.6 branches and 860.8 functions with it, improvements of 104.0\% and 80.7\%, respectively. The branch gain exceeds the function gain, which suggests that dependency-aware generation does more than reach additional entry points. Without dependency information, the LLM must guess how to construct the objects, call prerequisites, and satisfy parameter constraints that the sampled APIs require. Harnesses built on such guesses can compile yet still return early in shallow argument-checking paths. Supplying the dependencies lets the harness establish the required state first, so execution proceeds into the internal logic behind each API. 

\subsubsection{Dominance of Coverage Guidance}

\begin{figure*}[!t]
    \centering
    \includegraphics[width=0.97\linewidth]{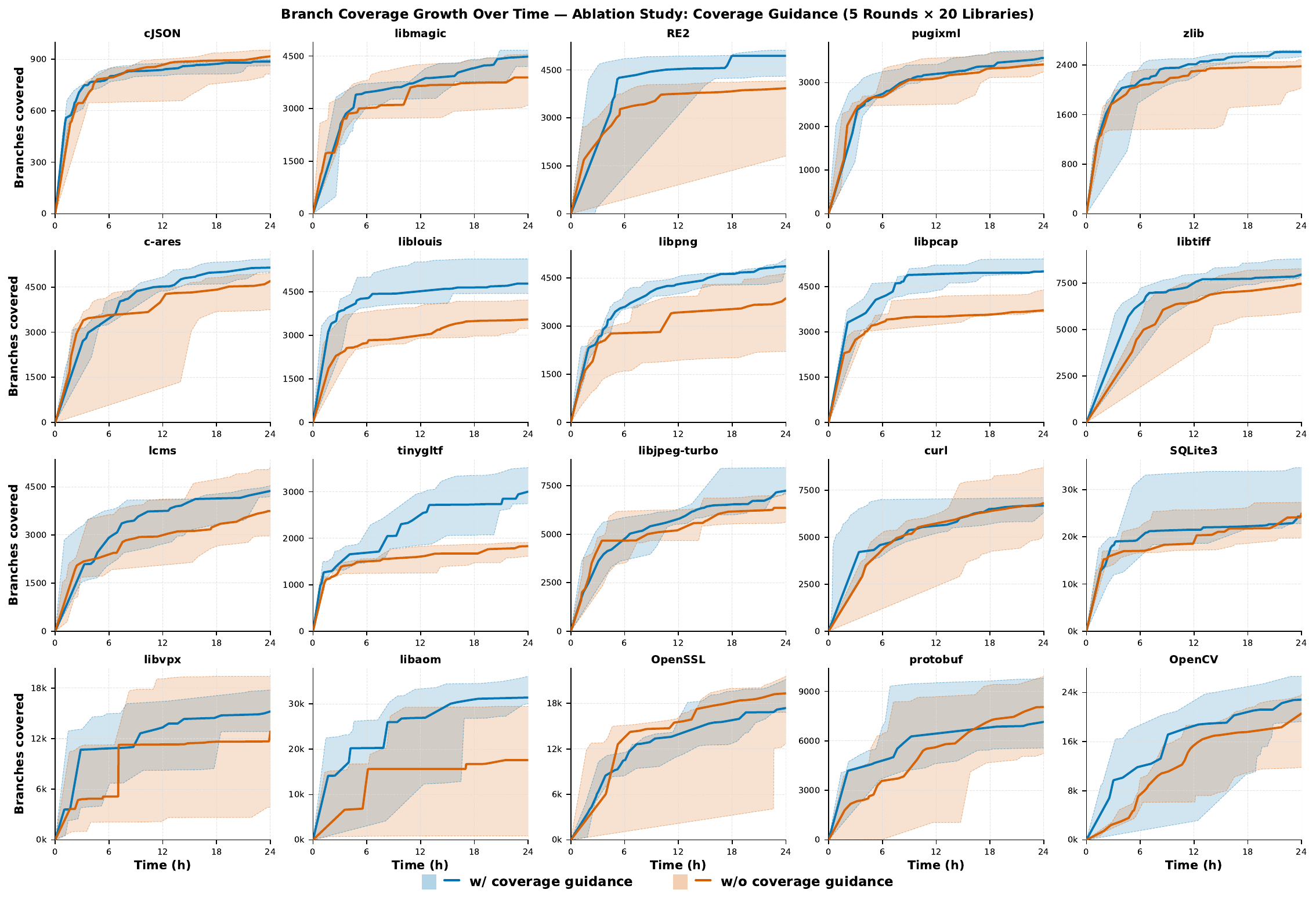}
    \caption{Effect of coverage guidance on branch-coverage growth over 24 hours for the 20 evaluated libraries. Lines show means over five independent runs, and shaded bands show 95\% t confidence intervals.}
    \label{fig:coverage_over_time}
\end{figure*}

The preceding comparisons associate actionable, multi-level feedback with higher harness productivity. To test whether coverage guidance itself contributes causally to this gain, we disable the \textit{Coverage Analyzer} while keeping the remainder of the workflow unchanged, and run both configurations on all 20 libraries for five independent 24-hour fuzzing trials. Removing the analyzer reduces aggregate branch coverage to \num{146508}, an 18.5\% decrease from the end-to-end setting. The per-library trajectories in \autoref{fig:coverage_over_time} show that this loss is not confined to a few outliers. With coverage guidance enabled, \sysname achieves obviously higher final coverage on 18 of the 20 libraries and steady coverage contribution on all libraries. Because the remaining workflow is unchanged, the degradation isolates the role of the feedback loop. The \textit{Coverage Analyzer} therefore does more than measure progress: it turns under-explored APIs, functions, and branches into scheduling decisions for subsequent harness generation, allowing the evolutionary workflow to sustain exploration instead of repeatedly revisiting already saturated behavior.

\subsection{Bug Detection}
\label{sec:bug_detection}

Across five independent 24-hour end-to-end runs on the 20 libraries, the \textit{Crash Analyzer} recorded \num{1098} raw crash reports. Call-stack de-duplication with \tool{CASR}~\cite{savidov2021casr}, followed by automated triage, reduced these reports to 128 candidate library bugs. The remaining reports were classified as duplicates or harness errors. Manual adjudication ultimately retained 102 of the 128 candidates (79.69\%) as genuine library bugs and rejected 26 as false positives. Internal reproduction eliminated 20 of these false positives: in 12 cases, the agent failed to retrieve the relevant API constraints from library documentation; in the other 8, it drew an incorrect root-cause conclusion from the available evidence. The remaining six candidates initially passed our internal review but were conservatively excluded after maintainer feedback (\autoref{sec:discuss}). As summarized in \extref{tab:bugs}{Table} (\apref{sec:detected-bugs}), we reported all 102 retained bugs upstream with a reproducer and root-cause analysis. To date, 90 have been acknowledged and fixed.

The 102 confirmed bugs span a broad range of defect classes, with integer overflows (32) and buffer overflows (22) being the two most common classes. To examine whether these findings translate into practical security impact, we analyzed the 21 bugs found in \tool{libaom}, the AV1 codec used by Chromium and other media software. Seven are directly reachable through \tool{libaom}'s own command-line tools, \tool{aomenc} and \tool{aomdec}, using crafted files or command-line arguments. Triggering them requires no custom harness. Their reachability through shipped applications provides independent evidence that they are not artifacts of fuzzing code, but defects on externally exposed program paths relevant to downstream consumers.

\begin{table}[t]
    \centering
    \setlength{\tabcolsep}{2.5pt}
    \begin{threeparttable}
        \caption{Crash-triage accuracy after manual adjudication.}
        \label{tab:crash_validation}
        \small
        \begin{tabular}{lrrrrrr}
            \toprule
            \textbf{System} & \textbf{Raw} & \textbf{Candidates} & \textbf{TP} & \textbf{FP} & \textbf{FP Rate} & \textbf{TP Rate} \\
            \midrule
            \tool{PromptFuzz} & \num{20737} & 89  & 5   & 84  & 94.38\%          & 5.62\% \\
            \tool{PromeFuzz}  & 6279 & 316 & 1   & 315 & 99.68\%          & 0.32\% \\
            \tool{ProteusFuzz} &\num{313437} & 439 & 4 & 435 & 99.09\% & 0.91\% \\
            \sysname          & 1098 & 128 & 102 & 26  & \textbf{20.31\%} & \textbf{79.69\%} \\
            \bottomrule
        \end{tabular}
        \begin{tablenotes}[flushleft]
            \item \textbf{Raw} denotes observed crashes before validation and deduplication and differs in each system. \textbf{Candidates} denote deduplicated crashes labeled as library bugs before manual adjudication.
        \end{tablenotes}
    \end{threeparttable}
\end{table}

\subsubsection{Comparison with Baselines}
To compare triage quality rather than the volume of crash manifestations, we evaluated all four systems over 25 independent 24-hour fuzzing trials across 5-round evaluations, and manually adjudicated the candidates that survived their respective pipelines. \autoref{tab:crash_validation} summarizes the results. \tool{PromptFuzz} rejected \num{20737} harnesses during validation, while its retained harnesses later yielded 89 candidates; only 5 were genuine library bugs, giving a false-positive rate of 94.38\%. \tool{PromeFuzz} produced \num{6279} raw crash reports and retained 316 candidates, only one of which was genuine, for a false-positive rate of 99.68\%. 
As an extension of \tool{PromeFuzz}, \tool{ProteusFuzz} shows a similar tendency, with a true-positive rate of only 0.91\%. 
By comparison, \sysname retained 128 candidates from \num{1098} raw reports and correctly identified 102 library bugs, limiting its false-positive rate to 20.31\%. All these bugs detected by other systems were also detectable by \sysname.

\tool{PromptFuzz} uses bounded execution as its validation oracle. It executes each generated harness and briefly fuzzes it, discarding the harness if either stage fails~\cite{ccs_24_promptfuzz}. Across our five generation runs, this procedure rejected \num{20737} harnesses. It is effective against shallow misuse that fails on common inputs or has a high trigger probability. Passing a bounded test, however, does not establish that every fuzz input preserves the library's semantic contract. Input-dependent violations of object state, ownership, or calculation can survive validation and emerge only during a longer campaign. 

\tool{PromeFuzz} goes beyond runtime filtering by asking an LLM to reason over the crash trace and statically retrieved context~\cite{ccs_25_promefuzz}. However, its one-step pipeline labeled \num{3568} of \num{6279} raw reports as potential bugs, and the deduplication left 316 candidates, of which 315 were harness errors (99.68\%). Our classification attributes 239 errors (75.8\%) to \emph{opaque object states}, where validity or ownership depends on runtime facts absent from the static harness and trace; and 55 (17.5\%) to \emph{misconstructed static context}, where trace-guided retrieval omits the constructor, helper API, or invariant that defines the applicable contract. Only the remaining 21 cases (6.7\%) are \emph{residual model errors} in which sufficient evidence was retrieved but misinterpreted. The first two evidence failures account for 93.3\% of the false positives and instantiate the observability problem in \autoref{sec:background:crash_validation_example}.


\subsubsection{Real-World Impact}

Eight of the 21 \tool{libaom} bugs are buffer overflows. One yields an arbitrary-write primitive that an attacker-controlled index escapes the intended bounds and determines the destination of an out-of-bounds store. Another yields an out-of-bounds read that discloses pointers from adjacent heap objects and can therefore defeat ASLR. Chaining these primitives produces a remote-code-execution path through AV1 encoding with the SVC feature enabled. This exploit chain demonstrates that \sysname reaches security-critical states and uncovers vulnerabilities with realistic attack surfaces. Four CVE IDs have been assigned to these vulnerabilities, and all have been fixed in the latest \tool{libaom} release. 


\begin{table}[!h]
    \centering
    \caption{\small{Branch coverage change from \tool{DeepSeek V3.2} to \tool{GLM-5.2.}}}
    \label{tab:llm_generalization}
    \fontsize{7.2pt}{8.0pt}\selectfont
    \setlength{\tabcolsep}{1pt}
    \sisetup{reset-text-series=false, text-series-to-math=true}
    \newcommand{\genlibhead}[1]{{\fontsize{6.8pt}{7.5pt}\selectfont #1}}
    \begin{tabular*}{\columnwidth}{@{\extracolsep{\fill}}lrrrrrrr@{}}
        \toprule
        \textbf{Model}
        & \genlibhead{cJSON}
        & \genlibhead{libmagic}
        & \genlibhead{RE2}
        & \genlibhead{pugixml}
        & \genlibhead{zlib}
        & \genlibhead{c-ares}
        & \genlibhead{liblouis} \\
        \midrule
        \tool{DeepSeek V3.2}
        & \num{880} & \num{4484} & \num{4069} & \num{3582} & \num{2598} & \num{5182} & \num{4911} \\
        \tool{GLM-5.2}
        & \textbf{\num{899}} & \textbf{\num{5381}} & \textbf{\num{4433}} & \textbf{\num{4076}} & \textbf{\num{2689}} & \textbf{\num{5781}} & \textbf{\num{5327}} \\
        \cmidrule(lr){1-8}
        $\Delta$
        & \num{2.16}\% & \num{20.00}\% & \num{8.95}\% & \num{13.79}\% & \num{3.50}\% & \num{11.56}\% & \num{8.47}\% \\
        \addlinespace[2pt]
        \toprule
        \textbf{Model}
        & \genlibhead{libpng}
        & \genlibhead{libpcap}
        & \genlibhead{libtiff}
        & \genlibhead{lcms}
        & \genlibhead{tinygltf}
        & \makecell[c]{\genlibhead{libjpeg}\\[-1pt]\genlibhead{-turbo}}
        & \genlibhead{curl} \\
        \midrule
        \tool{DeepSeek V3.2}
        & \num{4921} & \textbf{\num{5120}} & \num{8212} & \num{4537} & \num{3056} & \num{7541} & \num{6699} \\
        \tool{GLM-5.2}
        & \textbf{\num{5509}} & {\num{5030}} & \textbf{\num{10020}} & \textbf{\num{4866}} & \textbf{\num{3714}} & \textbf{\num{7852}} & \textbf{\num{8051}} \\
        \cmidrule(lr){1-8}
        $\Delta$
        & \num{11.95}\% & \num{-1.76}\% & \num{22.02}\% & \num{7.25}\% & \num{21.53}\% & \num{4.12}\% & \num{20.18}\% \\
        \addlinespace[2pt]
        \toprule
        \textbf{Model}
        & \genlibhead{libvpx}
        & \genlibhead{SQLite3}
        & \genlibhead{libaom}
        & \genlibhead{protobuf}
        & \genlibhead{OpenSSL}
        & \genlibhead{OpenCV}
        & \genlibhead{\textbf{Total}} \\
        \midrule
        \tool{DeepSeek V3.2}
        & \num{15680} & \num{26197} & \num{33308} & \num{7664} & \num{18227} & \num{12751} & \num{179619} \\
        \tool{GLM-5.2}
        & \textbf{\num{17532}} & \textbf{\num{31459}} & \textbf{\num{36702}} & \textbf{\num{12180}} & \textbf{\num{25879}} & \textbf{\num{14807}} & \textbf{\num{212187}} \\
        \cmidrule(lr){1-8}
        $\Delta$
        & \num{11.81}\% & \num{20.09}\% & \num{10.19}\% & \num{58.92}\% & \num{41.98}\% & \num{16.12}\% & \textbf{\num{18.13}}\% \\
        \bottomrule
    \end{tabular*}
\end{table}

\subsection{Generalization Across LLMs}
\label{sec:generalization}
Our main evaluation uses \tool{DeepSeek V3.2}, raising the question of whether \sysname's effectiveness depends on a particular LLM. To isolate this factor, we replaced the LLM backbone with \tool{GLM-5.2}~\cite{glm_52} while leaving the entire workflow unchanged. We then repeated the complete end-to-end experiment for five independent runs on all 20 libraries. \autoref{tab:llm_generalization} reports the mean branch coverage over these runs.

With \tool{GLM-5.2}, \sysname covers \num{212187} branches in total, an \num{18.13}\% improvement over the \num{179619} branches obtained with \tool{DeepSeek V3.2}. The improvement is broad rather than driven by a small subset of targets, on which coverage increases on 19 of the 20 libraries. The largest gains appear on \tool{protobuf} (\num{58.92}\%), \tool{OpenSSL} (\num{41.98}\%), and \tool{libtiff} (\num{22.02}\%). In particular, the substantial improvements on \tool{protobuf} and \tool{OpenSSL} suggest that stronger model capabilities can further benefit knowledge acquisition and strategy formation for complex libraries.
Model capability also affects crash reasoning. In a separate replay of the 20 false positives in \autoref{sec:bug_detection} that survived \sysname's automated triage, replacing \tool{DeepSeek V3.2} with \tool{GLM-5.2} correctly reclassified 14 as harness errors. This result confirms that residual triage errors can decrease as the underlying model improves. Together, these results show that a stronger backbone can improve both coverage and crash reasoning, while the ability to construct and evolve a working fuzzing campaign generalizes across the two model families as a property of \sysname's workflow.



\section{Discussion}
\label{sec:discuss}

\subsection{Statistical variance in library fuzzing}

Library fuzzing is stochastic, and LLM-based harness generation adds a second source of randomness on top of the fuzzer itself.
However, existing work (i.e., \tool{PromeFuzz}~\cite{ccs_25_promefuzz} and \tool{ProteusFuzz}~\cite{tdsc_26_proteusfuzz}) reports only a single harness-generation trial per library, which can produce misleading results. For example, \tool{PromeFuzz} reports 1.22$\times$ the branch coverage of \tool{PromptFuzz} on the same targets. In our evaluation in \autoref{sec:eval}, \tool{PromeFuzz} reaches only 0.95$\times$ the coverage of \tool{PromptFuzz} on the 15 libraries supported by both (\num{88266} versus \num{92909} branches) and is ahead on only 5 of them. We attribute the difference to the variance of LLM-generated harnesses.

To quantify both sources of randomness, we ran a nested experiment. For each LLM-based method and library, we performed five independent harness-generation trials, merged the valid harnesses of each trial into one composite harness, and ran five independent 24-hour fuzzing trials on it, yielding 25 runs per method--library pair. We measure dispersion with the coefficient of variation, $\mathrm{CV}=\sigma/\mu\times100\%$, which makes libraries of very different size comparable. Let $c_{i,j}$ be the branch coverage of the $j$-th fuzzing trial on the harness from the $i$-th generation trial. The harness CV is computed over the per-generation means $m_i=\frac{1}{5}\sum_{j}c_{i,j}$. The fuzzing CV is computed over $\{c_{i,1},\ldots,c_{i,5}\}$ within each generation trial and averaged over the five trials.

\autoref{tab:coverage-variance} shows that harness generation is the larger source of variation. Averaged over libraries, the harness CV exceeds the fuzzing CV for all three systems: 10.07\% versus 1.28\% for \sysname, 10.13\% versus 5.17\% for \tool{PromeFuzz}, and 12.96\% versus 2.63\% for \tool{PromptFuzz}. The same holds for every library for \sysname and \tool{PromptFuzz}, and for 13 of 17 libraries for \tool{PromeFuzz}. Repeating only the fuzzing phase controls fuzzer randomness but not LLM randomness. We therefore repeat both harness generation and the 24-hour fuzzing phase five times each and report the mean over all trials. To avoid distributional assumptions, we also apply the non-parametric Mann--Whitney U test (\apref{sec:mann-whitney-u-test-for-statistical-significance}). The per-library p-values show that \sysname's improvements are statistically significant for most library-baseline pairs.

\begin{table}[htbp]
    \centering
    \scriptsize
    \setlength{\tabcolsep}{4pt}
    \begin{threeparttable}
        \caption{Coefficient of variation across independent harness-generation and fuzzing trials.}
        \label{tab:coverage-variance}
        \begin{tabular}{lcccccc}
            \toprule
            \multirow{2}{*}{\textbf{Library}} & \multicolumn{2}{c}{\textbf{\sysname}} & \multicolumn{2}{c}{\textbf{\tool{PromeFuzz}}} & \multicolumn{2}{c}{\textbf{\tool{PromptFuzz}}} \\
            \cmidrule(lr){2-3} \cmidrule(lr){4-5} \cmidrule(lr){6-7}
             & \textbf{Harness} & \textbf{Fuzzing} & \textbf{Harness} & \textbf{Fuzzing} & \textbf{Harness} & \textbf{Fuzzing} \\
            \midrule
            cJSON         & 2.66\%  & 0.14\% & 5.01\%  & 3.50\%  & 2.25\%  & 0.05\% \\
            libmagic      & 5.57\%  & 2.37\% & 4.76\%  & 4.97\%  & 16.71\% & 3.70\% \\
            RE2           & 2.71\%  & 0.52\% & 1.59\%  & 4.59\%  & 8.59\%  & 2.94\% \\
            pugixml       & 2.73\%  & 0.42\% & 4.45\%  & 1.76\%  & N/A     & N/A \\
            zlib          & 2.04\%  & 0.40\% & 4.99\%  & 1.01\%  & 3.88\%  & 0.79\% \\
            c-ares        & 3.46\%  & 0.66\% & 4.30\%  & 2.67\%  & 4.87\%  & 0.50\% \\
            liblouis      & 10.36\% & 0.64\% & 11.48\% & 1.53\%  & 16.23\% & 8.31\% \\
            libpng        & 4.44\%  & 1.11\% & 31.09\% & 6.04\%  & 22.34\% & 1.97\% \\
            libpcap       & 4.02\%  & 2.21\% & 5.30\%  & 4.56\%  & 18.39\% & 2.50\% \\
            libtiff       & 2.80\%  & 1.27\% & 12.35\% & 11.46\% & 9.22\%  & 4.82\% \\
            lcms          & 9.25\%  & 1.38\% & 5.70\%  & 6.76\%  & 5.72\%  & 1.20\% \\
            tinygltf      & 7.61\%  & 2.58\% & 5.33\%  & 5.46\%  & N/A     & N/A \\
            libjpeg-turbo & 7.38\%  & 0.88\% & 16.98\% & 12.30\% & 36.23\% & 6.74\% \\
            curl          & 30.36\% & 1.59\% & 18.34\% & 1.72\%  & 6.90\%  & 1.24\% \\
            libvpx        & 24.28\% & 0.92\% & 23.54\% & 8.60\%  & 6.88\%  & 2.64\% \\
            SQLite3       & 13.60\% & 2.94\% & 4.49\%  & 4.34\%  & 29.31\% & 3.26\% \\
            libaom        & 7.13\%  & 0.92\% & 12.49\% & 6.69\%  & 6.96\%  & 0.87\% \\
            protobuf      & 15.03\% & 1.28\% & N/A     & N/A     & N/A     & N/A \\
            OpenSSL       & 28.46\% & 1.96\% & N/A     & N/A     & 12.85\% & 0.51\% \\
            OpenCV        & 17.47\% & 1.45\% & N/A     & N/A     & N/A     & N/A \\
            \midrule
            \textbf{Mean} & \textbf{10.07\%} & \textbf{1.28\%} & \textbf{10.13\%} & \textbf{5.17\%} & \textbf{12.96\%} & \textbf{2.63\%} \\
            \bottomrule
        \end{tabular}
        \begin{tablenotes}
            \item \textbf{Note:} CV denotes coefficient of variation. The harness columns compute CV over the mean coverage of 5 fuzzing trials for each of 5 independent harness-generation trials. The fuzzing columns compute the within-harness CV over 5 independent 24-hour fuzzing trials and then average the CV across the 5 generated harnesses. N/A indicates unsupported or unavailable settings.
        \end{tablenotes}
    \end{threeparttable}
\end{table}

\subsection{False Positives and Residual Error Modes}
Crash validation is essential to distinguish library defects from harness-induced failures. \autoref{tab:crash_validation} shows that \sysname's automated triage reduces the false-positive rate to 20.31\%, a substantial improvement over the 94.38\%, 99.68\%, and 99.09\% of \tool{PromptFuzz}, \tool{PromeFuzz}, and \tool{ProteusFuzz}, respectively. The remaining false positives fall into two groups, which we discuss below to give insights for further research.

The first group contains the 20 false positives eliminated by our own reproduction. They correspond to the two evidence stages of the \textit{Crash Analyzer}. In 12 cases, the agent failed to retrieve the API constraints that define valid usage from the library documentation, so it judged a harness-induced violation to be a library defect. In the other 8, the relevant evidence was available but the agent drew an incorrect root-cause conclusion. Both are limits of the current backbone and tool use rather than of the workflow, and the model-replacement study in \autoref{sec:generalization} points the same way: replaying these 20 cases with \tool{GLM-5.2} correctly reclassified 14 as harness errors.

The second group is harder, because internal evidence cannot remove it. The remaining 108 candidates passed our validation, yet maintainer feedback led us to exclude six (5.6\%) conservatively, leaving the final 102 bugs, of which 90 have been acknowledged or fixed to date. Three were considered expected behavior under undocumented contracts. For example, a \tool{libvpx} out-of-memory report was closed because the format permits resolutions up to $65536 \times 65536$, an expectation that the maintainers apply but that was not part of the contract available to the agent. Two were treated as robustness issues but deferred as out of scope for the current release. These are a disagreement about policy rather than correctness. The program does fail on the input, but the project does not regard the failure as a defect to fix. The last was a harness-side API misuse that escaped internal review, the same failure mode as the first group, and shows that validation can still error toward a plausible but wrong root cause.

These cases suggest two directions for crash analysis. Evidence-retrieval and reasoning errors should shrink with stronger models and better retrieval, as the replay above indicates. Undocumented contracts and project-specific expectations are different, because the information resides in maintainers' practice rather than in the library. A natural extension is to record maintainer decisions, such as accepted input-size limits and the classes of robustness reports a project declines, and supply them to the \textit{Crash Analyzer} as per-library context. Until then, we treat a manual check before disclosure as part of the workflow, and the 102 bugs we report are those that survived both that check and maintainer review.



\section{Related Work}
\label{sec:related}

\noindent\textbf{Automated library fuzzing.}
Library fuzzing has evolved from manual approaches to increasingly automated techniques. \tool{OSS-Fuzz}~\cite{oss_fuzz} pioneered large-scale library fuzzing but required substantial manual effort. Researchers have developed various approaches to automate harness generation.
Static analysis based approaches, like \tool{Fudge}~\cite{fudge}, \tool{FuzzGen}~\cite{fuzzgen}, \tool{GraphFuzz}~\cite{graphfuzz}, and \tool{AFGen}~\cite{afgen}, leverage static analysis to extract API usage patterns and generate fuzzing harnesses. \tool{Fudge} identifies API entry points, \tool{FuzzGen} extracts patterns from client applications, \tool{GraphFuzz} builds dataflow graphs for dependencies, and \tool{AFGen} focuses on whole-function fuzzing.
Dynamic analysis approaches, like \tool{APICraft}~\cite{apicraft} and \tool{Hopper}~\cite{hopper}, incorporate dynamic analysis to improve harness generation. \tool{APICraft} records API interactions during execution, while \tool{Hopper} uses interpretative execution to understand API behaviors.
Hybrid approaches, like \tool{RULF}~\cite{rulf} and \tool{UTopia}~\cite{utopia}, combine multiple techniques. \tool{RULF} traverses API dependency graphs to generate comprehensive harnesses for Rust libraries. \tool{UTopia} leverages existing unit tests to generate effective fuzz drivers.

\noindent\textbf{LLMs for fuzzing.}
The emergence of LLMs has opened new possibilities for automating and enhancing fuzzing processes. LLM-based harness generation, like \tool{OSS-Fuzz-Gen}~\cite{oss-fuzz-gen}, \tool{PromptFuzz} ~\cite{ccs_24_promptfuzz} and \tool{PromeFuzz}~\cite{ccs_25_promefuzz}, pioneered using LLMs to generate fuzzing harnesses from API documentation and code. \tool{ProteusFuzz}~\cite{tdsc_26_proteusfuzz} extends \tool{PromeFuzz} with an agent that activates at coverage plateaus, selects a stagnation target with a control-flow heuristic, and revises seeds or harnesses, using an API call chain from a precomputed dependency graph for state-related blockers. It is the closest to our work in using coverage feedback to repair harnesses. \sysname differs in three respects: its diagnosis combines coverage at five levels instead of a single stagnation line, its harness generation recovers four kinds of API dependencies instead of a bare call chain, and it also automates campaign construction and validates crashes with runtime object states. \tool{TitanFuzz}~\cite{titan_fuzz} specifically targets deep learning libraries, showing LLMs can handle complex, domain-specific APIs. LLM-enhanced fuzzing components, like \tool{CodaMosa}~\cite{CodaMosa}, use LLMs to overcome coverage plateaus in test generation. \tool{Universal Fuzzing}~\cite{xia2023universal} demonstrates fuzzing across diverse domains without domain-specific customization. \tool{Xu et al.}~\cite{xu2025directedgreyboxfuzzinglarge} explore using LLMs for directed greybox fuzzing toward specific targets or code regions.


\section{Conclusion}
\label{sec:conclusion}

We presented \sysname, a multi-agent system that automates library fuzzing from a source tree and evolves the campaign from concrete execution evidence. Specialized interfaces let agents construct and repair every lifecycle artifact, hierarchical coverage diagnosis turns a coverage gap into an exploration target, dependency-aware generation realizes that target as a harness with its required API lifecycle, and runtime object states ground the verdict on each crash.
Across 20 real-world C/C++ libraries, \sysname completed all end-to-end rounds without human intervention and covered 45.1\%--191.2\% more branches than \tool{OSS-Fuzz}, \tool{OSS-Fuzz-Gen}, \tool{PromptFuzz}, \tool{PromeFuzz}, and \tool{ProteusFuzz}. \sysname also found 102 library bugs, of which upstream maintainers have acknowledged or fixed 90.



\bibliographystyle{IEEEtran}
\bibliography{references}

@misc{afl,
    title = {American fuzzy lop},
    author = {Michal Zalewski},
    howpublished = "\url{http://lcamtuf.coredump.cx/afl/}",
    year = {Accessed 2026}
}

@misc{fuzz_introspector,
  title={Fuzz Introspector -- introspect, extend and optimise fuzzers},
  author={Open Source Security Foundation (OpenSSF)},
  year={Accessed 2022},
  url={https://github.com/ossf/fuzz-introspector}
}

@misc{source_based_code_coverage,
  title={Source-based Code Coverage},
  author={LLVM},
  url={https://clang.llvm.org/docs/SourceBasedCodeCoverage.html},
  year = {Accessed 2026}
}

@misc{llvm-cov-tool,
  title={llvm-cov - emit coverage information},
  author={LLVM},
  url={https://llvm.org/docs/CommandGuide/llvm-cov.html},
  year={Accessed 2026}
}

@misc{tree_sitter,
  title={An incremental parsing system for programming tools},
  author={Max Brunsfeld and the Tree-Sitter contributors},
  url={https://tree-sitter.github.io/tree-sitter/}
}

@misc{tree_reduce,
  title={treereduce},
  author={Langston Barrett},
  url={https://github.com/langston-barrett/treereduce}
}

@inproceedings{savidov2021casr,
  title = {{{Casr-Cluster}}: Crash Clustering for Linux Applications},
  author = {Savidov, Georgy and Fedotov, Andrey},
  booktitle = {2021 Ivannikov ISPRAS Open Conference (ISPRAS)},
  pages = {47--51},
  year = {2021},
  organization = {IEEE},
  doi = {10.1109/ISPRAS53967.2021.00012},
}

@misc{oss-fuzz-gen,
    title = {oss-fuzz-gen},
    author = {Google},
    howpublished = "\url{https://github.com/google/oss-fuzz-gen}",
    year={Accessed 2026}
}

@misc{fuzzer_split,
    title = {How To Split A Fuzzer-Generated Input Into Several},
    author = {Google},
    howpublished = "\url{https://github.com/google/fuzzing/blob/master/docs/split-inputs.md}"
}

@conference{oss_fuzz,
  title={{OSS-Fuzz}-Google's continuous fuzzing service for open source software},
  booktitle={Proceedings of the 26th USENIX Conference on Security Symposium (technical sessions)},
  author={Serebryany, Kostya},
  publisher = {USENIX Association},
  year={2017}
}

@misc{ubsan,
    title = "Undefined Behavior Sanitizer - Official Documentation",
    author = {LLVM},
    howpublished = "\url{https://clang.llvm.org/docs/UndefinedBehaviorSanitizer.html}",
    year = {Accessed 2026}
}

@misc{deepseek_model,
    title = {Hugging Face: DeepSeek V3.2 Model},
    author = {DeepSeek-AI},
    howpublished = "\url{https://huggingface.co/deepseek-ai/DeepSeek-V3.2}",
    year={Accessed 2026}
}

@misc{promefuzz_issue,
    title = {Can {PromeFuzz} be used to fuzz {OpenSSL}?},
    author = {{PromeFuzz developers}},
    howpublished = "\url{https://github.com/pvz122/PromeFuzz/issues/8}",
    year={Accessed 2026}
}

@misc{promptfuzz_author_response,
    title = {{PromptFuzz} author response to its official release},
    author = {{PromptFuzz developers}},
    howpublished = "\url{https://github.com/FuzzAnything/PromptFuzz/releases/tag/v1.0.0}",
    year={Accessed 2026}
}

@inproceedings{aflfast,
    author = {B\"{o}hme, Marcel and Pham, Van-Thuan and Roychoudhury, Abhik},
    title = {Coverage-Based Greybox Fuzzing as Markov Chain},
    year = {2016},
    booktitle = {Proceedings of the 2016 ACM SIGSAC Conference on Computer and Communications Security},
    pages = {1032–1043},
    numpages = {12},
}

@inproceedings {aflpp,
    author = {Andrea Fioraldi and Dominik Maier and Heiko Ei{\ss}feldt and Marc Heuse},
    title = {{AFL++} : Combining Incremental Steps of Fuzzing Research},
    booktitle = {14th USENIX Workshop on Offensive Technologies (WOOT 20)},
    year = {2020},
}

@inproceedings{fudge,
  title={Fudge: fuzz driver generation at scale},
  author={Babi{\'c}, Domagoj and Bucur, Stefan and Chen, Yaohui and Ivan{\v{c}}i{\'c}, Franjo and King, Tim and Kusano, Markus and Lemieux, Caroline and Szekeres, L{\'a}szl{\'o} and Wang, Wei},
  booktitle={Proceedings of the 2019 27th ACM Joint Meeting on European Software Engineering Conference and Symposium on the Foundations of Software Engineering},
  pages={975--985},
  year={2019}
}

@inproceedings{fuzzgen,
  title={{FuzzGen}: Automatic Fuzzer Generation},
  author={Ispoglou, Kyriakos and Austin, Daniel and Mohan, Vishwath and Payer, Mathias},
  booktitle={29th USENIX Security Symposium (USENIX Security 20)},
  pages={2271--2287},
  year={2020}
}

@INPROCEEDINGS {afgen,
    author = {Y. Liu and Y. Wang and T. Bao and X. Jia and Z. Zhang and P. Su},
    booktitle = {2024 IEEE Symposium on Security and Privacy (SP)},
    title = {AFGen: Whole-Function Fuzzing for Applications and Libraries},
    year = {2024},
    pages = {11-11},
}

@inproceedings{TitanFuzz,
    author = {Deng, Yinlin and Xia, Chunqiu Steven and Peng, Haoran and Yang, Chenyuan and Zhang, Lingming},
    title = {Large Language Models Are Zero-Shot Fuzzers: Fuzzing Deep-Learning Libraries via Large Language Models},
    year = {2023},
    booktitle = {Proceedings of the 32nd ACM SIGSOFT International Symposium on Software Testing and Analysis},
    pages = {423–435},
}

@inproceedings{intelligen,
  title={IntelliGen: Automatic driver synthesis for fuzz testing},
  author={Zhang, Mingrui and Liu, Jianzhong and Ma, Fuchen and Zhang, Huafeng and Jiang, Yu},
  booktitle={2021 IEEE/ACM 43rd International Conference on Software Engineering: Software Engineering in Practice (ICSE-SEIP)},
  pages={318--327},
  year={2021},
  organization={IEEE}
}

@inproceedings{apicraft,
  title={{APICraft}: Fuzz Driver Generation for Closed-source {SDK} Libraries},
  author={Zhang, Cen and Lin, Xingwei and Li, Yuekang and Xue, Yinxing and Xie, Jundong and Chen, Hongxu and Ying, Xinlei and Wang, Jiashui and Liu, Yang},
  booktitle={30th USENIX Security Symposium (USENIX Security 21)},
  pages={2811--2828},
  year={2021}
}

@inproceedings{graphfuzz,
  author={Green, Harrison and Avgerinos, Thanassis},
  booktitle={2022 IEEE/ACM 44th International Conference on Software Engineering (ICSE)}, 
  title={GraphFuzz: Library API Fuzzing with Lifetime-aware Dataflow Graphs}, 
  year={2022},
  pages={1070-1081},
}

@inproceedings{rulf,
  title={RULF: Rust library fuzzing via API dependency graph traversal},
  author={Jiang, Jianfeng and Xu, Hui and Zhou, Yangfan},
  booktitle={2021 36th IEEE/ACM International Conference on Automated Software Engineering (ASE)},
  pages={581--592},
  year={2021},
  organization={IEEE}
}

@inproceedings{utopia,
  title={UTOPIA: Automatic Generation of Fuzz Driver using Unit Tests},
  author={Jeong, Bokdeuk and Jang, Joonun and Yi, Hayoon and Moon, Jiin and Kim, Junsik and Jeon, Intae and Kim, Taesoo and Shim, WooChul and Hwang, Yong Ho},
  booktitle={2023 IEEE Symposium on Security and Privacy (SP)},
  pages={746--762},
  year={2022},
  organization={IEEE Computer Society}
}

@inproceedings{hopper,
  author = {Chen, Peng and Xie, Yuxuan and Lyu, Yunlong and Wang, Yuxiao and Chen, Hao},
  title = {HOPPER: Interpretative Fuzzing for Libraries},
  booktitle = {ACM Conference on Computer and Communications Security (CCS)},
  date = {2023-11-26/2023-11-30},
  year={2023},
  address = {Copenhagen, Denmark},
}

@inproceedings{asan,
    author = {Serebryany, Konstantin and Bruening, Derek and Potapenko, Alexander and Vyukov, Dmitry},
    title = {AddressSanitizer: A Fast Address Sanity Checker},
    year = {2012},
    publisher = {USENIX Association},
    booktitle = {Proceedings of the 2012 USENIX Conference on Annual Technical Conference},
    pages = {28},
    numpages = {1},
    location = {Boston, MA},
    series = {USENIX ATC'12}
}

@inproceedings{titan_fuzz,
  title={Large Language Models are Zero-Shot Fuzzers: Fuzzing Deep-Learning Libraries via Large Language Models},
  author={Deng, Yinlin and Xia, Chunqiu Steven and Peng, Haoran and Yang, Chenyuan and Zhang, Lingming},
  booktitle={Proceedings of the 32nd ACM SIGSOFT International Symposium on Software Testing and Analysis},
  pages={423--435},
  year={2023}
}

@article{xia2023universal,
      title={Universal Fuzzing via Large Language Models}, 
      author={Chunqiu Steven Xia and Matteo Paltenghi and Jia Le Tian and Michael Pradel and Lingming Zhang},
      year={2023},
      eprint={2308.04748},
      archivePrefix={arXiv},
      journal={arXiv preprint arXiv:2308.04748},
      primaryClass={cs.SE}
}

@INPROCEEDINGS{CodaMosa,
  author={Lemieux, Caroline and Inala, Jeevana Priya and Lahiri, Shuvendu K. and Sen, Siddhartha},
  booktitle={2023 IEEE/ACM 45th International Conference on Software Engineering (ICSE)}, 
  title={CodaMosa: Escaping Coverage Plateaus in Test Generation with Pre-trained Large Language Models}, 
  year={2023},
  pages={919-931},
}

@inproceedings{yang2024sweagent,
  title={{SWE}-agent: Agent-Computer Interfaces Enable Automated Software Engineering},
  author={John Yang and Carlos E Jimenez and Alexander Wettig and Kilian Lieret and Shunyu Yao and Karthik R Narasimhan and Ofir Press},
  booktitle={The Thirty-eighth Annual Conference on Neural Information Processing Systems},
  year={2024},
  url={https://arxiv.org/abs/2405.15793}
}

@article{TOSEM_2023_human_side_fuzzing,
    author = {Nourry, Olivier and Kashiwa, Yutaro and Lin, Bin and Bavota, Gabriele and Lanza, Michele and Kamei, Yasutaka},
    title = {The Human Side of Fuzzing: Challenges Faced by Developers during Fuzzing Activities},
    year = {2023},
    issue_date = {January 2024},
    publisher = {Association for Computing Machinery},
    address = {New York, NY, USA},
    volume = {33},
    number = {1},
    issn = {1049-331X},
    url = {https://doi.org/10.1145/3611668},
    doi = {10.1145/3611668},
    journal = {ACM Trans. Softw. Eng. Methodol.},
    month = nov,
    articleno = {14},
    numpages = {26}
}

@ARTICLE{fuzzing_challenges_reflections,
  author={Boehme, Marcel and Cadar, Cristian and ROYCHOUDHURY, Abhik},
  journal={IEEE Software}, 
  title={Fuzzing: Challenges and Reflections}, 
  year={2021},
  volume={38},
  number={3},
  pages={79-86},
  doi={10.1109/MS.2020.3016773}
  }

@misc{oss-fuzz-guide,
    title = {OSS-Fuzz Guide: Setting up a new project},
    howpublished = {\url{https://google.github.io/oss-fuzz/getting-started/new-project-guide/}},
    year = {Accessed 2026}
  }

@article{fuzzing_vulnerability_discovery_survey,
title = {Fuzzing vulnerability discovery techniques: Survey, challenges and future directions},
journal = {Computers \& Security},
volume = {120},
pages = {102813},
year = {2022},
issn = {0167-4048},
doi = {https://doi.org/10.1016/j.cose.2022.102813},
url = {https://www.sciencedirect.com/science/article/pii/S0167404822002073},
author = {Craig Beaman and Michael Redbourne and J. Darren Mummery and Saqib Hakak},
}

@article{fuzzers_for_stateful_systems,
author = {Daniele, Cristian and Andarzian, Seyed Behnam and Poll, Erik},
title = {Fuzzers for Stateful Systems: Survey and Research Directions},
year = {2024},
issue_date = {September 2024},
publisher = {Association for Computing Machinery},
address = {New York, NY, USA},
volume = {56},
number = {9},
issn = {0360-0300},
url = {https://doi.org/10.1145/3648468},
doi = {10.1145/3648468},
journal = {ACM Comput. Surv.},
month = apr,
articleno = {222},
numpages = {23}
}

@misc{xu2025directedgreyboxfuzzinglarge,
      title={Directed Greybox Fuzzing via Large Language Model}, 
      author={Hanxiang Xu and Yanjie Zhao and Haoyu Wang},
      year={2025},
      eprint={2505.03425},
      archivePrefix={arXiv},
      primaryClass={cs.CR},
      url={https://arxiv.org/abs/2505.03425}, 
}

@inproceedings{ccs_25_fuzzer_usability,
    author = {Zhao, Yunze and Guo, Wentao and Goldstein, Harrison and Votipka, Daniel and Fulton, Kelsey R. and Mazurek, Michelle L.},
    title = {A Qualitative Analysis of Fuzzer Usability and Challenges},
    year = {2025},
    isbn = {9798400715259},
    publisher = {Association for Computing Machinery},
    address = {New York, NY, USA},
    url = {https://doi.org/10.1145/3719027.3765055},
    doi = {10.1145/3719027.3765055},
    booktitle = {Proceedings of the 2025 ACM SIGSAC Conference on Computer and Communications Security},
    pages = {2504–2518},
    numpages = {15},
    location = {Taipei, Taiwan},
    series = {CCS '25}
}

@inproceedings{soups_21_clang_libfuzzer,
    author = {Pl\"{o}ger, Stephan and Meier, Mischa and Smith, Matthew},
    title = {A qualitative usability evaluation of the clang static analyzer and libFuzzer with CS students and CTF players},
    year = {2021},
    isbn = {978-1-939133-25-0},
    publisher = {USENIX Association},
    address = {USA},
    booktitle = {Proceedings of the Seventeenth USENIX Conference on Usable Privacy and Security},
    articleno = {29},
    numpages = {20},
    series = {SOUPS'21}
}

@inproceedings{chi_2023_fuzzer_usability,
    author = {Pl\"{o}ger, Stephan and Meier, Mischa and Smith, Matthew},
    title = {A Usability Evaluation of AFL and libFuzzer with CS Students},
    year = {2023},
    isbn = {9781450394215},
    publisher = {Association for Computing Machinery},
    address = {New York, NY, USA},
    url = {https://doi.org/10.1145/3544548.3581178},
    doi = {10.1145/3544548.3581178},
    booktitle = {Proceedings of the 2023 CHI Conference on Human Factors in Computing Systems},
    articleno = {186},
    numpages = {18},
    location = {Hamburg, Germany},
    series = {CHI '23}
}

@ARTICLE{tase_2021_fuzzing_survey,
  author = {Valentin J. M. Man{\`{e}}s and HyungSeok Han and Choongwoo Han and Sang Kil Cha and Manuel Egele and Edward J. Schwartz and Maverick Woo},
  title = {The Art, Science, and Engineering of Fuzzing: A Survey},
  journal = {IEEE Transactions on Software Engineering},
  volume = {47},
  number = {11},
  pages = {2312--2331},
  year = 2021
}

@inproceedings {sec_2020_symcc,
  author = {Sebastian Poeplau and Aur{\'e}lien Francillon},
  title = {Symbolic execution with {SymCC}: Don{\textquoteright}t interpret, compile!},
  booktitle = {29th USENIX Security Symposium (USENIX Security 20)},
  year = {2020},
  isbn = {978-1-939133-17-5},
  pages = {181--198},
  url = {https://www.usenix.org/conference/usenixsecurity20/presentation/poeplau},
  publisher = {USENIX Association},
  month = aug
}

@inproceedings {sec_2023_dynsql,
  author = {Zu-Ming Jiang and Jia-Ju Bai and Zhendong Su},
  title = {{DynSQL}: Stateful Fuzzing for Database Management Systems with Complex and Valid {SQL} Query Generation},
  booktitle = {32nd USENIX Security Symposium (USENIX Security 23)},
  year = {2023},
  isbn = {978-1-939133-37-3},
  address = {Anaheim, CA},
  pages = {4949--4965},
  url = {https://www.usenix.org/conference/usenixsecurity23/presentation/jiang-zu-ming},
  publisher = {USENIX Association},
  month = aug
}

@inproceedings{fuzzing_2023_fuzz_blockers,
  author = {Gao, Wentao and Pham, Van-Thuan and Liu, Dongge and Chang, Oliver and Murray, Toby and Rubinstein, Benjamin I.P.},
  title = {Beyond the Coverage Plateau: A Comprehensive Study of Fuzz Blockers (Registered Report)},
  year = {2023},
  isbn = {9798400702471},
  publisher = {Association for Computing Machinery},
  address = {New York, NY, USA},
  url = {https://doi.org/10.1145/3605157.3605177},
  doi = {10.1145/3605157.3605177},
  booktitle = {Proceedings of the 2nd International Fuzzing Workshop},
  pages = {47–55},
  numpages = {9},
  location = {Seattle, WA, USA},
  series = {FUZZING 2023}
}

@inproceedings{ccs_24_promptfuzz,
  author = {Lyu, Yunlong and Xie, Yuxuan and Chen, Peng and Chen, Hao},
  title = {Prompt Fuzzing for Fuzz Driver Generation},
  year = {2024},
  isbn = {9798400706363},
  publisher = {Association for Computing Machinery},
  address = {New York, NY, USA},
  url = {https://doi.org/10.1145/3658644.3670396},
  doi = {10.1145/3658644.3670396},
  booktitle = {Proceedings of the 2024 on ACM SIGSAC Conference on Computer and Communications Security},
  pages = {3793–3807},
  numpages = {15},
  location = {Salt Lake City, UT, USA},
  series = {CCS '24}
}

@inproceedings{ccs_25_promefuzz,
  author = {Liu, Yuwei and Deng, Junquan and Jia, Xiangkun and Wang, Yanhao and Wang, Minghua and Huang, Lin and Wei, Tao and Su, Purui},
  title = {PromeFuzz: A Knowledge-Driven Approach to Fuzzing Harness Generation with Large Language Models},
  year = {2025},
  isbn = {9798400715259},
  publisher = {Association for Computing Machinery},
  address = {New York, NY, USA},
  url = {https://doi.org/10.1145/3719027.3765222},
  doi = {10.1145/3719027.3765222},
  booktitle = {Proceedings of the 2025 ACM SIGSAC Conference on Computer and Communications Security},
  pages = {1559–1573},
  numpages = {15},
  location = {Taipei, Taiwan},
  series = {CCS '25}
}

@inproceedings{issta_21_seed_selection,
  author = {Herrera, Adrian and Gunadi, Hendra and Magrath, Shane and Norrish, Michael and Payer, Mathias and Hosking, Antony L.},
  title = {Seed selection for successful fuzzing},
  year = {2021},
  isbn = {9781450384599},
  publisher = {Association for Computing Machinery},
  address = {New York, NY, USA},
  url = {https://doi.org/10.1145/3460319.3464795},
  doi = {10.1145/3460319.3464795},
  booktitle = {Proceedings of the 30th ACM SIGSOFT International Symposium on Software Testing and Analysis},
  pages = {230–243},
  numpages = {14},
  location = {Virtual, Denmark},
  series = {ISSTA 2021}
}

@misc{openai_practical_guide_building_agents,
  author = {OpenAI},
  title = {A practical guide to building agents},
  howpublished = {\url{https://cdn.openai.com/business-guides-and-resources/a-practical-guide-to-building-agents.pdf}},
  year = {Accessed 2026}
}

@misc{modelcontext_writing_effective_tools,
  author = {ModelContextProtocol team.},
  title = {Writing Effective Tools for Agents},
  howpublished = {\url{https://modelcontextprotocol.info/docs/tutorials/writing-effective-tools/}},
  year = {Accessed 2026}
}

@misc{gdb_batch_model,
  author = {Free Software Foundation, Inc.},
  title={GDB Non-interactive Batch Mode},
  howpublished = {\url{https://www.sourceware.org/gdb/current/onlinedocs/gdb.html/Mode-Options.html}},
  year={Accessed 2026}
}

@misc{claude_reminder,
  title={Agent design lessons from Claude Code},
  author={Janne},
  howpublished = {\url{https://jannesklaas.github.io/ai/2025/07/20/claude-code-agent-design.html}},
  year={Accessed 2026}
}

@misc{github_api,
  title = {GitHub REST API},
  howpublished = {\url{https://docs.github.com/en/rest?apiVersion=2022-11-28}},
  year = {Accessed 2026}
}

@misc{deepseek_v32,
      title={DeepSeek-V3.2: Pushing the Frontier of Open Large Language Models}, 
      author={{DeepSeek-AI} and Liu, Aixin and Mei, Aoxue and others},
      year={2025},
      eprint={2512.02556},
      archivePrefix={arXiv},
      primaryClass={cs.CL},
      url={https://arxiv.org/abs/2512.02556}, 
}

@inproceedings{vllm,
  author = {Kwon, Woosuk and Li, Zhuohan and Zhuang, Siyuan and Sheng, Ying and Zheng, Lianmin and Yu, Cody Hao and Gonzalez, Joseph and Zhang, Hao and Stoica, Ion},
  title = {Efficient Memory Management for Large Language Model Serving with PagedAttention},
  year = {2023},
  isbn = {9798400702297},
  publisher = {Association for Computing Machinery},
  address = {New York, NY, USA},
  url = {https://doi.org/10.1145/3600006.3613165},
  doi = {10.1145/3600006.3613165},
  booktitle = {Proceedings of the 29th Symposium on Operating Systems Principles},
  pages = {611–626},
  numpages = {16},
  location = {Koblenz, Germany},
  series = {SOSP '23}
}

@inproceedings{evaluating_fuzz_testing,
  author = {Klees, George and Ruef, Andrew and Cooper, Benji and Wei, Shiyi and Hicks, Michael},
  title = {Evaluating Fuzz Testing},
  year = {2018},
  isbn = {9781450356930},
  publisher = {Association for Computing Machinery},
  address = {New York, NY, USA},
  url = {https://doi.org/10.1145/3243734.3243804},
  doi = {10.1145/3243734.3243804},
  booktitle = {Proceedings of the 2018 ACM SIGSAC Conference on Computer and Communications Security},
  pages = {2123–2138},
  numpages = {16},
  location = {Toronto, Canada},
  series = {CCS '18}
}

@article{fuzzing_build_failures,
  author = {Nourry, Olivier and Kashiwa, Yutaro and Shang, Weiyi and Shu, Honglin and Kamei, Yasutaka},
  title = {My Fuzzers Won’t Build: An Empirical Study of Fuzzing Build Failures},
  year = {2025},
  issue_date = {February 2025},
  publisher = {Association for Computing Machinery},
  address = {New York, NY, USA},
  volume = {34},
  number = {2},
  issn = {1049-331X},
  url = {https://doi.org/10.1145/3688842},
  doi = {10.1145/3688842},
  journal = {ACM Trans. Softw. Eng. Methodol.},
  month = jan,
  articleno = {29},
  numpages = {30},
}

@misc{glm_52,
  author = {{Z.ai}},
  title = {{GLM-5.2: Built for Long-Horizon Tasks}},
  howpublished = {\url{https://z.ai/blog/glm-5.2}},
  year = {2026}
}

@misc{zlib_manual,
  author = {Jean-loup Gailly and Mark Adler},
  title = {{zlib 1.3.1 Manual}},
  howpublished = {\url{https://www.zlib.net/manual.html}},
  year = {2024},
  note = {Accessed August 2026}
}

@misc{libaom_image_header,
  author = {Aomdeia Organization},
  title = {aom\_image.h},
  howpublished = {\url{https://aomedia.googlesource.com/aom/+/refs/heads/main/aom/aom_image.h}},
  year = {2026},
  note = {Accessed August 2026}
}

@inproceedings {usenix25_g2fuzz,
  author = {Kunpeng Zhang and Zongjie Li and Daoyuan Wu and Shuai Wang and Xin Xia},
  title = {{Low-Cost} and Comprehensive Non-textual Input Fuzzing with {LLM-Synthesized} Input Generators},
  booktitle = {34th USENIX Security Symposium (USENIX Security 25)},
  year = {2025},
  isbn = {978-1-939133-52-6},
  address = {Seattle, WA},
  pages = {6999--7018},
  url = {https://www.usenix.org/conference/usenixsecurity25/presentation/zhang-kunpeng},
  publisher = {USENIX Association},
  month = aug
}

@article{tdsc_26_proteusfuzz,
  author={Wei, Jin and Liu, Yuwei and Qian, Kunwei and Shi, Hongmei and Wang, Changhui and Chen, Ping and Wang, Yanhao and Wang, Yi and Mao, Bing and Dai, Jun and Sun, Xiaoyan},
  journal={ IEEE Transactions on Dependable and Secure Computing },
  title={{ ProteusFuzz: From Static Harness Generation to Adaptive, Agent-Guided Library Fuzzing }},
  year={5555},
  volume={},
  number={01},
  ISSN={1941-0018},
  pages={1-17},
  doi={10.1109/TDSC.2026.3730860},
  url = {https://doi.ieeecomputersociety.org/10.1109/TDSC.2026.3730860},
  publisher={IEEE Computer Society},
  address={Los Alamitos, CA, USA},
  month=sep
}

@IEEEtranBSTCTL{IEEEexample:BSTcontrol,
  CTLdash_repeated_names = "no"
}

\ifdefined\bodyonly\else
\clearpage
\appendices


\section{System-specific Evaluation Settings}
\label{sec:system-specific-setting}

To ensure a controlled and reproducible experimental environment, we self-hosted \tool{DeepSeek V3.2}~\cite{deepseek_v32} with \tool{vLLM}~\cite{vllm} for all LLM-dependent components, thereby avoiding variability caused by external API rate limits or service-side changes. Following the official DeepSeek recommendations~\cite{deepseek_model}, we configured inference with a temperature of 1.0 and a top-$p$ value of 0.95. We used \tool{AFL++}~\cite{aflpp} as the fuzzing engine in all experiments and measured coverage with Source-based Code Coverage instrumentation~\cite{source_based_code_coverage} and \tool{llvm-cov}~\cite{llvm-cov-tool}. We excluded coverage from third-party libraries (e.g., \tool{absl} for \tool{RE2}) so that the reported results reflect only the target library. We enabled the \tool{ASAN} and \tool{UBSAN} sanitizers in all fuzzing campaigns to detect memory-safety and undefined-behaviour bugs. For the bounded execution policy in \autoref{eq:coverage_growth_rate}, we set the default time budget to one hour, the plateau threshold $\epsilon$ to 0.01\%, and the window $w$ to one minute.

To compare \sysname with the baseline systems under a common evaluation protocol, we allocated up to 24 hours to harness generation and an additional 24 hours to standalone fuzzing. For \tool{OSS-Fuzz}, whose harnesses had already been prepared by developers, we fuzzed each existing harness for 24 hours on a dedicated CPU core. For the LLM-based systems, we followed the evaluation protocol of \tool{PromeFuzz}~\cite{ccs_25_promefuzz}. The four generation-based baselines---\tool{OSS-Fuzz-Gen}, \tool{PromptFuzz}, \tool{PromeFuzz} and \tool{ProteusFuzz}---used five parallel LLM workers to generate harnesses for up to 24 hours, whereas \sysname used a single LLM worker to execute its complete end-to-end workflow for 24 hours. We then merged the harnesses retained by each system into a composite harness and fuzzed it for an additional 24 hours.

\tool{OSS-Fuzz-Gen} was the only exception: we compiled and fuzzed each of its generated harnesses independently for 24 hours because its harness-specific build scripts prevented the harnesses from being merged. Consistent with its original evaluation, \tool{PromeFuzz} performed codebase summarization with 50 parallel LLM workers. We evaluated \tool{OSS-Fuzz-Gen} over the complete API set rather than its default subset. \tool{ProteusFuzz} is an extension of \tool{PromeFuzz} and follows the same evaluation protocol of \tool{PromeFuzz}. All baseline campaigns used the dictionaries and seed corpora collected by \tool{OSS-Fuzz}. In contrast, \sysname received neither, allowing us to assess its ability to prepare these auxiliary inputs autonomously.

As shown in \autoref{sec:discuss}, coverage can vary substantially across independent runs because of randomness in both fuzzing and LLM-based harness generation, with harness generation contributing the dominant source of variation. Following best practices for fuzzing evaluation~\cite{evaluating_fuzz_testing}, we repeated each generation-based system for five independent end-to-end rounds. Within each round, we repeated the subsequent 24-hour standalone fuzzing campaign five times, yielding 25 fuzzing trials for each system--library pair. We report the mean across these trials throughout the evaluation. Because \sysname embeds fuzzing within its evolutionary workflow, \autoref{tab:eval_overview} also reports the branch coverage accumulated during its five 24-hour end-to-end runs, without the additional standalone fuzzing phase. We denote this result as \textit{FuzzAgent (E2E)}.

\section{Internal Analysis of Automation}
\label{sec:automation-details}

Beyond the end-to-end completion and input utility evaluated in \autoref{sec:automation-efficiency}, we examine how \sysname realizes this automation internally. We analyze whether campaign operations are grounded in domain-specific interfaces and how agent activity is distributed across campaign setup, harness evolution, fuzzing, and feedback analysis.

\subsection{Tool-Grounded Domain Automation}

General coding agents typically expose file operations and a shell, leaving the model to reconstruct instrumentation, fuzzing, coverage extraction, and debugging from low-level commands. \sysname retains these primitives but adds high-level interfaces that enable controlled strategies, structured domain analysis and actionable refinement.

\begin{table}[t]
    \centering
    \footnotesize
    \setlength{\tabcolsep}{2.0pt}
    \begin{threeparttable}
        \caption{Average tool invocations in five end-to-end rounds.}
        \label{tab:instr-count}
        \begin{tabular*}{\columnwidth}{@{\extracolsep{\fill}}lrrrrrrr@{}}
            \toprule
            & \multicolumn{2}{c}{\textit{General Tools}} & \multicolumn{4}{c}{\textcolor{RoyalBlue}{\textbf{FuzzAgent Interfaces}}} & \\
            \cmidrule(lr){2-3}\cmidrule(lr){4-7}
            \textbf{Agent} & \textbf{File} & \textbf{Bash} & \textcolor{RoyalBlue}{\textbf{Web}} & \textcolor{RoyalBlue}{\textbf{Fuzz}} & \textcolor{RoyalBlue}{\textbf{Cov.}} & \textcolor{RoyalBlue}{\textbf{Crash}} & \textbf{Total} \\
            \midrule
            Build    & 178   & 207  & 52  & 49   & 0    & 0    & 485   \\
            Dict.    & 146   & 177  & 370 & 0    & 0    & 0    & 693   \\
            Seed     & 19    & 563  & 416 & 0    & 0    & 0    & 998   \\
            Harness  & 11343 & 1962 & 60  & 1125 & 0    & 0    & 14490 \\
            Fuzzer   & 1552  & 3386 & 0   & 612  & 0    & 0    & 5550  \\
            Coverage & 562   & 330  & 0   & 0    & 3282 & 0    & 4174  \\
            Crash    & 1913  & 982  & 0   & 0    & 0    & 1213 & 4108  \\
            \midrule
            \textbf{Total} & \textbf{15714} & \textbf{7605} & \textbf{898} & \textbf{1786} & \textbf{3282} & \textbf{1213} & \textbf{30498} \\
            \bottomrule
            \end{tabular*}
        \begin{tablenotes}[flushleft]
            \item \textbf{File} and \textbf{Bash} are generic coding-agent tools. \textbf{Web}, \textbf{Fuzz} (build/compile/fuzz), \textbf{Cov.}, and \textbf{Crash} are interfaces implemented by \sysname; the latter three encode library-fuzzing-specific operations and verification.
        \end{tablenotes}
    \end{threeparttable}
\end{table}

\autoref{tab:instr-count} reports \num{30498} invocations averaged in five end-to-end round. File navigation is the largest category (51.52\%), reflecting the need to recover context from a continuously evolving workspace, while generic Bash accounts for 24.94\%. The custom interfaces account for the remaining \num{7179} calls (23.54\%). Once routine file navigation is excluded, 48.56\% of all operations use these custom interfaces rather than Bash. These interfaces form \sysname's execution layer that encapsules core workflows and domain functionality rather than convenience wrappers of general-purpose fuzzing. This division of labor lets agents focus on semantic diagnosis, strategy, and adaptation instead of reconstructing these mechanisms through fragile Bash sequences. The invocation distribution thus shows that \sysname operationalizes expertise through actively used domain interfaces, distinguishing it from a general coding agent that relies primarily on files and Bash.

\subsection{Agent-Time Allocation}

We further measure how agent activity is distributed during each 24-hour execution. This distribution reveals whether automation is dominated by one-time environment setup or sustains the iterative generation, execution, and analysis required to evolve a fuzzing campaign.

\begin{table}[t]
    \centering
    \scriptsize
    \setlength{\tabcolsep}{2.7pt}
    \begin{threeparttable}
        \caption{Agent-time distribution across the 24-hour end-to-end execution of \sysname.}
        \label{tab:agent-ratio}
        \begin{tabular}{lrrrrrrr}
            \toprule
            \textbf{Library} & \textbf{Builder} & \textbf{Dict.} & \textbf{Seed} & \textbf{Harness} & \textbf{Fuzzer} & \textbf{Crash} & \textbf{Coverage} \\
            \midrule
            cJSON         & 0.32\%  & 1.00\% & 1.33\% & 36.92\% & 37.96\% & 10.56\% & 11.91\% \\
            libmagic      & 8.26\%  & 1.33\% & 1.38\% & 34.26\% & 38.87\% & 6.16\%  & 9.74\%  \\
            RE2           & 11.38\% & 0.80\% & 1.18\% & 21.81\% & 47.81\% & 6.13\%  & 10.90\% \\
            pugixml       & 1.22\%  & 1.79\% & 1.82\% & 37.03\% & 41.16\% & 6.80\%  & 10.16\% \\
            zlib          & 1.52\%  & 1.15\% & 1.71\% & 27.22\% & 42.02\% & 14.02\% & 12.35\% \\
            c-ares        & 2.05\%  & 1.46\% & 2.16\% & 33.43\% & 40.06\% & 8.98\%  & 11.87\% \\
            liblouis      & 2.06\%  & 0.70\% & 0.99\% & 45.31\% & 26.66\% & 13.56\% & 10.72\% \\
            libpng        & 0.59\%  & 0.78\% & 1.47\% & 32.41\% & 32.90\% & 20.49\% & 11.36\% \\
            libpcap       & 4.05\%  & 1.00\% & 1.11\% & 35.81\% & 34.40\% & 9.98\%  & 13.65\% \\
            libtiff       & 13.85\% & 0.85\% & 1.31\% & 27.28\% & 28.83\% & 16.98\% & 10.91\% \\
            lcms          & 4.26\%  & 0.83\% & 1.04\% & 29.10\% & 26.69\% & 22.21\% & 15.87\% \\
            tinygltf      & 1.21\%  & 0.78\% & 0.74\% & 39.76\% & 33.90\% & 11.79\% & 11.82\% \\
            libjpeg-turbo & 4.68\%  & 0.81\% & 1.11\% & 37.03\% & 27.99\% & 19.64\% & 8.74\%  \\
            curl          & 5.59\%  & 1.28\% & 0.89\% & 27.27\% & 34.97\% & 13.77\% & 16.23\% \\
            libvpx        & 0.76\%  & 0.60\% & 1.29\% & 43.65\% & 19.22\% & 17.13\% & 17.35\% \\
            SQLite3       & 4.29\%  & 0.70\% & 1.64\% & 43.29\% & 30.76\% & 9.88\%  & 9.45\%  \\
            libaom        & 3.75\%  & 0.86\% & 1.46\% & 34.03\% & 34.64\% & 8.13\%  & 17.12\% \\
            protobuf      & 6.55\%  & 0.99\% & 1.24\% & 33.05\% & 26.14\% & 19.60\% & 12.44\% \\
            OpenSSL       & 3.52\%  & 1.07\% & 1.60\% & 28.11\% & 27.90\% & 13.04\% & 24.77\% \\
            OpenCV        & 2.92\%  & 0.74\% & 1.39\% & 34.19\% & 28.78\% & 13.26\% & 18.72\% \\
            \midrule
            \textbf{Average} & \textbf{4.14\%} & \textbf{0.98\%} & \textbf{1.34\%} & \textbf{34.05\%} & \textbf{33.08\%} & \textbf{13.11\%} & \textbf{13.30\%} \\
            \bottomrule
        \end{tabular}
        \begin{tablenotes}[flushleft]
            \item \textbf{Note:} Percentages denote the fraction of agent activity time within a 24-hour execution. Rows sum to 100\% up to rounding.
        \end{tablenotes}
    \end{threeparttable}
\end{table}

As shown in \autoref{tab:agent-ratio}, the three setup agents---the \textit{Library Builder}, \textit{Dictionary Generator}, and \textit{Seed Generator}---together account for only 6.46\% of agent activity time on average. The remaining 93.54\% is devoted to evolving and evaluating the campaign after setup. Harness generation (34.05\%) and fuzzer execution (33.08\%) form the largest components, while crash and coverage analysis contribute another 26.41\% in total. Thus, the workflow does not spend most of its effort merely constructing an executable starting point. It continually invests comparable effort in producing harnesses, exercising them, and interpreting the resulting feedback.

The allocation also varies with the target. Builder activity ranges from 0.32\% on \tool{cJSON} to 13.85\% on \tool{libtiff}, while crash and coverage analysis together range from 15.90\% on \tool{libmagic} to 38.08\% on \tool{lcms}. This variation reflects the stateful scheduler: once setup succeeds, activity shifts among generation, execution, and analysis according to the evidence produced by the current campaign rather than following a fixed per-agent schedule.

\section{Mann-Whitney U Test for Statistical Significance}
\label{sec:mann-whitney-u-test-for-statistical-significance}

Because coverage values are not guaranteed to follow a normal distribution, we use the non-parametric Mann--Whitney U test to assess whether the observed coverage differences between \sysname and each baseline are statistically significant. \autoref{tab:mann-whitney} reports the per-library p-values. These results complement the mean coverage numbers in \autoref{tab:eval_overview} by showing that \sysname's improvements are statistically significant for most available library-baseline pairs. In the four pairs where a baseline is significantly higher, the baseline's advantage is confined to a single library: \tool{SQLite3} and \tool{OpenSSL} for \tool{PromptFuzz}, \tool{curl} for \tool{ProteusFuzz}, and \tool{OpenSSL} for \tool{OSS-Fuzz}. Against \tool{ProteusFuzz}, the differences on \tool{cJSON} and \tool{SQLite3} are not significant.

\begin{table}[htbp]
	\centering
	\scriptsize
	\setlength{\tabcolsep}{3pt}
	\begin{threeparttable}
		\caption{Mann--Whitney U-test p-values comparing \sysname with baseline fuzzers.}
		\label{tab:mann-whitney}
		\begin{tabular}{lccccc}
			\toprule
			\textbf{Library} & \textbf{\makecell{Prome-\\Fuzz}} & \textbf{\makecell{Prompt-\\Fuzz}} & \textbf{\makecell{Proteus-\\Fuzz}} & \textbf{\makecell{OSS-\\Fuzz}} & \textbf{\makecell{OSS-Fuzz-\\Gen}} \\
			\midrule
			cJSON         & 1.37E-08 & 2.04E-08 & 0.090 & 8.85E-11 & 8.85E-11 \\
			libmagic      & 2.37E-09 & 0.133 & 1.29E-09 & 9.73E-11 & 9.73E-11 \\
			RE2           & 1.41E-09 & 1.41E-09 & 1.29E-09 & 9.73E-11 & 9.73E-11 \\
			pugixml       & 1.39E-09 & N/A & 1.27E-09 & 9.49E-11 & 9.49E-11 \\
			zlib          & 1.39E-09 & 2.25E-09 & 1.28E-09 & 9.56E-11 & 9.56E-11 \\
			c-ares        & 4.61E-05 & 0.197 & 1.29E-09 & 9.73E-11 & 9.73E-11 \\
			liblouis      & 1.39E-09 & 2.16E-03 & 1.27E-09 & 0.198 & 9.54E-11 \\
			libpng        & 1.41E-09 & 1.41E-09 & 1.29E-09 & 9.69E-11 & 9.69E-11 \\
			libpcap       & 1.41E-09 & 1.41E-09 & 1.29E-09 & 9.71E-11 & 9.71E-11 \\
			libtiff       & 1.84E-08 & 1.42E-09 & 1.29E-09 & 9.73E-11 & 9.73E-11 \\
			lcms          & 2.55E-09 & 1.16E-06 & 1.28E-09 & 9.61E-11 & 9.61E-11 \\
			tinygltf      & 1.41E-09 & N/A & 5.77E-07 & 9.71E-11 & 9.71E-11 \\
			libjpeg-turbo & 1.40E-09 & 1.40E-09 & 1.28E-09 & 4.48E-03 & 9.61E-11 \\
			curl          & 7.41E-03 & 4.45E-04 & 1.29E-09 & 1.05E-04 & 9.72E-11 \\
			libvpx        & 1.39E-09 & 1.39E-09 & 1.27E-09 & 9.50E-11 & 9.50E-11 \\
			SQLite3       & 3.02E-07 & 0.020 & 0.228 & 0.198 & 0.071 \\
			libaom        & 1.42E-09 & 1.41E-09 & 1.29E-09 & 9.73E-11 & 9.73E-11 \\
			protobuf      & N/A & N/A & N/A & 9.73E-11 & 9.73E-11 \\
			OpenSSL       & N/A & 3.90E-05 & N/A & 9.73E-11 & 9.73E-11 \\
			OpenCV        & N/A & N/A & N/A & 9.73E-11 & 9.73E-11 \\
			\midrule
			\textbf{$p<0.05$} & \textbf{17} & \textbf{14} & \textbf{15} & \textbf{18} & \textbf{19} \\
			Baseline higher & 0 & 2 & 1 & 1 & 0 \\
			\bottomrule
		\end{tabular}
		\begin{tablenotes}
			\item \textbf{Note:} Each entry reports the p-value of a two-sided Mann--Whitney U test comparing the coverage distribution of \sysname against the corresponding baseline on the same library. N/A indicates unavailable baseline results. Values below 0.01 are in scientific notation; the others are rounded to three decimals. The $p<0.05$ row counts libraries with a significant difference in either direction, and the next row counts those where the baseline's coverage is higher: \tool{SQLite3} and \tool{OpenSSL} for \tool{PromptFuzz}, \tool{curl} for \tool{ProteusFuzz}, and \tool{OpenSSL} for \tool{OSS-Fuzz}.
		\end{tablenotes}
	\end{threeparttable}
\end{table}

\balance
\providecommand{\yes}{{\color{ForestGreen}\ding{51}}}
\providecommand{\no}{{\color{black!35}\ding{55}}}
\providecommand{\rep}{R}
\providecommand{\cnf}{C}
\providecommand{\iss}[2]{{\footnotesize\href{#1}{\##2}}}

\onecolumn
\section{Detected Library Bugs}\label{sec:detected-bugs}

\begingroup
\small
\setlength{\tabcolsep}{1pt}
\setlength{\LTcapwidth}{\textwidth}
\renewcommand{\arraystretch}{1.05}
\begin{longtable}{@{}r l@{\hspace{4pt}} l l@{\hspace{3pt}} l c c c c c c@{}}
    \caption{Library bugs discovered by \sysname and detection coverage of baselines (\yes\ detected, \no\ missed). \textbf{Issue}: upstream issue ID. \textbf{St.}: \rep\ = reported, \cnf\ = confirmed/fixed. ``--'' = not evaluated.}
    \label{tab:bugs} \\
    \toprule
    \textbf{ID} & \textbf{Library} & \textbf{Crash Location} & \textbf{Vuln. Type} & \textbf{Issue} & \textbf{St.} & \textbf{OSS-Fuzz} & \textbf{\makecell{OSS-Fuzz-\\Gen}} & \textbf{\makecell{Prompt-\\Fuzz}} & \textbf{\makecell{Prome-\\Fuzz}} & \textbf{\makecell{Proteus-\\Fuzz}} \\
    \midrule
    \endfirsthead
    \multicolumn{11}{l}{\small\itshape Table~\ref{tab:bugs} continued from previous page}\\
    \toprule
    \textbf{ID} & \textbf{Library} & \textbf{Crash Location} & \textbf{Vuln. Type} & \textbf{Issue} & \textbf{St.} & \textbf{OSS-Fuzz} & \textbf{\makecell{OSS-Fuzz-\\Gen}} & \textbf{\makecell{Prompt-\\Fuzz}} & \textbf{\makecell{Prome-\\Fuzz}} & \textbf{\makecell{Proteus-\\Fuzz}} \\
    \midrule
    \endhead
    \midrule
    \multicolumn{11}{r}{\small\itshape continued on next page}\\
    \endfoot
    \bottomrule
    \endlastfoot
    1   & cJSON         & cJSON.c:2524                         & Integer Overflow  & Embargo & \rep & \no & \no & \no & \no & \no \\
    2   & cJSON         & cJSON.c:2001                         & Use After Free    & Embargo & \rep & \no & \no & \no & \no & \no \\
    3   & cJSON         & cJSON.c:1963                         & Type Mismatch     & Embargo & \rep & \no & \no & \yes & \no & \no \\
    4   & libmagic      & softmagic.c:771                      & Null Pointer      & \iss{https://bugs.astron.com/view.php?id=763}{763} & \cnf & \no & \no & \no & \no & \no \\
    5   & libmagic      & encoding.c:286                       & Segment Violation & \iss{https://bugs.astron.com/view.php?id=762}{762} & \cnf & \no & \no & \no & \no & \no \\
    6   & pugixml       & pugixml.cpp:5207                     & Buffer Overflow   & \iss{https://github.com/zeux/pugixml/issues/714}{714} & \cnf & \no & \no & \no & \no & \no \\
    7   & pugixml       & pugixml.cpp:232                      & Segment Violation & \iss{https://github.com/zeux/pugixml/issues/713}{713} & \cnf & \no & \no & \no & \no & \no \\
    8   & pugixml       & pugixml.cpp:4565                     & Memory Alignment  & Embargo & \rep & \no & \no & \no & \no & \no \\
    9   & zlib          & gzlib.c:389                          & Integer Overflow  & \iss{https://github.com/madler/zlib/issues/1222}{1222} & \cnf & \no & \no & \no & \no & \no \\
    10  & c-ares        & ares\_process.c:1156                 & Integer Overflow  & \iss{https://github.com/c-ares/c-ares/issues/1144}{1144} & \cnf & \no & \no & \no & \no & \no \\
    11  & liblouis      & pattern.c:801                        & Segment Violation & \iss{https://github.com/liblouis/liblouis/issues/1987}{1987} & \cnf & \no & \no & \yes & \no & \no \\
    12  & liblouis      & compileTranslationTable.c:430        & Segment Violation & \iss{https://github.com/liblouis/liblouis/issues/1978}{1978} & \cnf & \no & \no & \no & \no & \no \\
    13  & liblouis      & compileTranslationTable.c:296        & Segment Violation & \iss{https://github.com/liblouis/liblouis/issues/1976}{1976} & \cnf & \no & \no & \no & \no & \no \\
    14  & liblouis      & compileTranslationTable.c:4865       & Memory Leak       & \iss{https://github.com/liblouis/liblouis/issues/1986}{1986} & \cnf & \no & \no & \no & \no & \no \\
    15  & liblouis      & lou\_translateString.c:1343          & Buffer Overflow   & \iss{https://github.com/liblouis/liblouis/issues/1982}{1982} & \cnf & \no & \no & \no & \no & \no \\
    16  & liblouis      & logging.c:57                         & Buffer Overflow   & \iss{https://github.com/liblouis/liblouis/issues/1980}{1980} & \cnf & \no & \no & \no & \no & \no \\
    17  & liblouis      & lou\_backTranslateString.c:257       & Buffer Overflow   & \iss{https://github.com/liblouis/liblouis/issues/1984}{1984} & \cnf & \no & \no & \no & \no & \no \\
    18  & liblouis      & lou\_translateString.c:1343          & Buffer Overflow   & {\footnotesize by email} & \cnf & \no & \no & \no & \no & \no \\
    19  & libpng        & pngwutil.c:2182                      & Buffer Overflow   & Embargo & \rep & \no & \no & \no & \no & \no \\
    20  & libpng        & pngrtran.c:1070                      & Integer Overflow  & Embargo & \rep & \no & \no & \no & \no & \no \\
    21  & libpcap       & pcap-util.c:567                      & Memory Alignment  & \iss{https://github.com/the-tcpdump-group/libpcap/issues/1634}{1634} & \cnf & \no & \no & \no & \yes & \no \\
    22  & libpcap       & bpf\_filter.c:112                    & Abort             & {\footnotesize by email} & \cnf & \no & \no & \no & \no & \no \\
    23  & libtiff       & tif\_luv.c:1318                      & Integer Overflow  & \iss{https://gitlab.com/libtiff/libtiff/-/work_items/850}{850} & \cnf & \no & \no & \no & \no & \yes \\
    24  & lcms          & cmsio0.c:1530                        & Documentation     & \iss{https://github.com/mm2/Little-CMS/issues/546}{546} & \cnf & \no & \no & \no & \no & \no \\
    25  & lcms          & cmsio0.c:1606                        & Documentation     & \iss{https://github.com/mm2/Little-CMS/issues/545}{545} & \cnf & \no & \no & \no & \no & \no \\
    26  & lcms          & cmsnamed.c:808                       & Null Pointer      & \iss{https://github.com/mm2/Little-CMS/issues/544}{544} & \cnf & \no & \no & \no & \no & \yes \\
    27  & tinygltf      & stb\_image\_write.h:1263             & Integer Overflow  & Embargo & \rep & \no & \no & \no & \no & \no \\
    28  & tinygltf      & stb\_image.h:1210                    & Integer Overflow  & Embargo & \rep & \no & \no & \no & \no & \no \\
    29  & libjpeg-turbo & turbojpeg.c:937                      & Type Mismatch     & Embargo & \rep & \no & \no & \yes & \no & \no \\
    30  & libjpeg-turbo & jcarith.c:438                        & Type Mismatch     & Embargo & \rep & \no & \no & \no & \no & \no \\
    31  & libjpeg-turbo & cmyk.h:55                            & Type Mismatch     & Embargo & \rep & \no & \no & \no & \no & \no \\
    32  & libvpx        & vpx\_image.c:263                     & Null Pointer      & \iss{https://issues.webmproject.org/issues/496881245}{496881245} & \cnf & \no & \no & \yes & \no & \no \\
    33  & libvpx        & firstpass.c:678                      & Integer Overflow  & \iss{https://issues.webmproject.org/u/1/issues/496881243}{496881243} & \cnf & \no & \no & \no & \no & \no \\
    34  & libvpx        & bitwriter\_buffer.c:48               & Integer Overflow  & \iss{https://issues.webmproject.org/u/1/issues/496807676}{496807676} & \cnf & \no & \no & \no & \no & \no \\
    35  & libvpx        & encodeframe.c:831                    & Integer Overflow  & \iss{https://issues.webmproject.org/u/1/issues/505215075}{505215075} & \cnf & \no & \no & \no & \no & \no \\
    36  & libvpx        & vpx\_dsp\_common.h:89                & Integer Overflow  & \iss{https://issues.webmproject.org/u/1/issues/505208797}{505208797} & \cnf & \no & \no & \no & \no & \no \\
    37  & libvpx        & vp9\_cx\_iface.c:349                 & Integer Overflow  & \iss{https://issues.webmproject.org/u/1/issues/505902439}{505902439} & \cnf & \no & \no & \no & \no & \no \\
    38  & libvpx        & vp9\_svc\_layercontext.c:468         & Integer Overflow  & \iss{https://issues.webmproject.org/u/1/issues/505947426}{505947426} & \cnf & \no & \no & \no & \no & \no \\
    39  & libvpx        & vp9\_ratectrl.c:2172                 & Integer Overflow  & \iss{https://issues.webmproject.org/u/1/issues/505623919}{505623919} & \cnf & \no & \no & \no & \no & \no \\
    40  & libvpx        & vp9\_svc\_layercontext.c:1334        & Integer Overflow  & \iss{https://issues.webmproject.org/u/1/issues/505631147}{505631147} & \cnf & \no & \no & \no & \no & \no \\
    41  & libvpx        & vp9\_cx\_iface.c:501                 & Integer Overflow  & \iss{https://issues.webmproject.org/u/1/issues/505286902}{505286902} & \cnf & \no & \no & \yes & \no & \no \\
    42  & libvpx        & vp9\_encoder.h:1079                  & Integer Overflow  & \iss{https://issues.webmproject.org/u/1/issues/505286908}{505286908} & \cnf & \no & \no & \no & \no & \no \\
    43  & libvpx        & vp9\_quantize.c:324                  & Buffer Overflow   & \iss{https://issues.webmproject.org/u/1/issues/505286900}{505286900} & \cnf & \no & \no & \no & \no & \no \\
    44  & libvpx        & onyx\_if.c:970                       & Buffer Overflow   & \iss{https://issues.webmproject.org/u/1/issues/505902433}{505902433} & \cnf & \no & \no & \no & \no & \no \\
    45  & libvpx        & vp9\_encodeframe.c:580               & Buffer Overflow   & \iss{https://issues.webmproject.org/issues/442161280}{442161280} & \cnf & \no & \no & \no & \no & \no \\
    46  & libvpx        & sad4d\_avx512.c:35                   & Buffer Overflow   & \iss{https://issues.webmproject.org/u/1/issues/505665613}{505665613} & \cnf & \no & \no & \no & \no & \no \\
    47  & libvpx        & vp9\_bitstream.c:59                  & Buffer Overflow   & \iss{https://issues.webmproject.org/issues/441668134}{441668134} & \cnf & \no & \no & \no & \no & \no \\
    48  & libvpx        & vp9\_rd.c:351                        & Buffer Overflow   & \iss{https://issues.webmproject.org/u/1/issues/505947421}{505947421} & \cnf & \no & \no & \no & \no & \no \\
    49  & libvpx        & vp9\_decoder.c:311                   & Segment Violation & \iss{https://issues.webmproject.org/u/1/issues/505665625}{505665625} & \cnf & \no & \no & \no & \no & \no \\
    50  & libvpx        & highbd\_sad\_avx2.c:29               & Segment Violation & \iss{https://issues.webmproject.org/u/1/issues/505286906}{505286906} & \cnf & \no & \no & \no & \no & \no \\
    51  & libvpx        & vp9\_svc\_layercontext.c:1007        & Segment Violation & \iss{https://issues.webmproject.org/u/1/issues/505830937}{505830937} & \cnf & \no & \no & \no & \no & \no \\
    52  & libvpx        & highbd\_variance\_impl\_sse2.asm:217 & Segment Violation & \iss{https://issues.webmproject.org/u/1/issues/505902436}{505902436} & \cnf & \no & \no & \no & \no & \no \\
    53  & libvpx        & variance.c:269                       & Segment Violation & \iss{https://issues.webmproject.org/u/1/issues/505947415}{505947415} & \cnf & \no & \no & \no & \no & \no \\
    54  & libvpx        & vp9\_encoder.c:4104                  & Segment Violation & \iss{https://issues.webmproject.org/u/1/issues/505286909}{505286909} & \cnf & \no & \no & \no & \no & \yes \\
    55  & libvpx        & onyxd\_if.c:126                      & Segment Violation & \iss{https://issues.webmproject.org/u/1/issues/505286904}{505286904} & \cnf & \no & \no & \no & \no & \no \\
    56  & libvpx        & vp9\_bitstream.c:402                 & Segment Violation & \iss{https://issues.webmproject.org/issues/442105459}{442105459} & \cnf & \no & \no & \no & \no & \no \\
    57  & libvpx        & vp8\_dx\_iface.c:134                 & Assertion Fail    & \iss{https://issues.webmproject.org/u/1/issues/505947417}{505947417} & \cnf & \no & \no & \no & \no & \no \\
    58  & libvpx        & onyx\_if.c:1233                      & Assertion Fail    & \iss{https://issues.webmproject.org/u/1/issues/506017558}{506017558} & \cnf & \no & \no & \no & \no & \no \\
    59  & libvpx        & vp9\_encodeframe.c:5471              & Assertion Fail    & \iss{https://issues.webmproject.org/u/1/issues/505665619}{505665619} & \cnf & \no & \no & \no & \no & \no \\
    60  & sqlite        & sqlite3.c:263698                     & Integer Overflow  & Embargo & \rep & \no & \no & \no & \no & \no \\
    61  & libaom        & firstpass.c:830                      & Integer Overflow  & \iss{https://aomedia.issues.chromium.org/u/1/issues/502933723}{502933723} & \cnf & \no & \no & \no & \no & \no \\
    62  & libaom        & pass2\_strategy.c:308                & Integer Overflow  & \iss{https://aomedia.issues.chromium.org/u/1/issues/503013394}{503013394} & \cnf & \no & \no & \no & \no & \no \\
    63  & libaom        & intra\_mode\_search\_utils.h:107     & Integer Overflow  & \iss{https://aomedia.issues.chromium.org/u/1/issues/503171644}{503171644} & \cnf & \no & \no & \no & \no & \no \\
    64  & libaom        & var\_based\_part.c:419               & Integer Overflow  & \iss{https://aomedia.issues.chromium.org/u/1/issues/503187935}{503187935} & \cnf & \no & \no & \no & \no & \no \\
    65  & libaom        & ratectrl.c:508                       & Integer Overflow  & \iss{https://aomedia.issues.chromium.org/u/1/issues/503197490}{503197490} & \cnf & \no & \no & \no & \no & \no \\
    66  & libaom        & variance.c:280                       & Integer Overflow  & \iss{https://aomedia.issues.chromium.org/u/1/issues/503689686}{503689686} & \cnf & \no & \no & \no & \no & \no \\
    67  & libaom        & noise\_model.c:1270                  & Integer Overflow  & \iss{https://aomedia.issues.chromium.org/u/1/issues/503810640}{503810640} & \cnf & \no & \no & \no & \no & \no \\
    68  & libaom        & psnr.c:65                            & Integer Overflow  & \iss{https://aomedia.issues.chromium.org/u/1/issues/503689686}{503689686} & \cnf & \no & \no & \no & \no & \no \\
    69  & libaom        & encodetxb.c:615                      & Buffer Overflow   & \iss{https://aomedia.issues.chromium.org/issues/471095598}{471095598} & \cnf & \no & \no & \no & \no & \no \\
    70  & libaom        & av1\_dx\_iface.c:1330                & Buffer Overflow   & \iss{https://aomedia.issues.chromium.org/u/1/issues/503197473}{503197473} & \cnf & \no & \no & \no & \no & \no \\
    71  & libaom        & svc\_layercontext.c:444              & Buffer Overflow   & \iss{https://aomedia.issues.chromium.org/issues/503975732}{503975732} & \cnf & \no & \no & \no & \no & \no \\
    72  & libaom        & pass2\_strategy.c:1215               & Buffer Overflow   & \iss{https://aomedia.issues.chromium.org/issues/504317456}{504317456} & \cnf & \no & \no & \no & \no & \no \\
    73  & libaom        & svc\_layercontext.c:315              & Buffer Overflow   & \iss{https://aomedia.issues.chromium.org/issues/503993985}{503993985} & \cnf & \no & \no & \no & \no & \no \\
    74  & libaom        & aq\_cyclicrefresh.c:237              & Buffer Overflow   & \iss{https://aomedia.issues.chromium.org/issues/503993984}{503993984} & \cnf & \no & \no & \no & \no & \no \\
    75  & libaom        & encodetxb.c:617                      & Buffer Overflow   & \iss{https://aomedia.issues.chromium.org/u/1/issues/503171639}{503171639} & \cnf & \no & \no & \no & \no & \no \\
    76  & libaom        & av1\_fwd\_txfm2d\_avx2.c:1450        & Buffer Overflow   & \iss{https://aomedia.issues.chromium.org/issues/503689685}{503689685} & \cnf & \no & \no & \no & \no & \no \\
    77  & libaom        & av1\_ext\_ratectrl.c:186             & Segment Violation & \iss{https://aomedia.issues.chromium.org/u/1/issues/503326506}{503326506} & \cnf & \no & \no & \no & \no & \no \\
    78  & libaom        & av1\_dx\_iface.c:953                 & Segment Violation & \iss{https://aomedia.issues.chromium.org/issues/503691210}{503691210} & \cnf & \no & \no & \no & \no & \yes \\
    79  & libaom        & intra\_mode\_search.c:358            & Assertion Fail    & \iss{https://aomedia.issues.chromium.org/u/1/issues/503993976}{503993976} & \cnf & \no & \no & \no & \no & \no \\
    80  & libaom        & encoder.c:2314                       & Assertion Fail    & \iss{https://aomedia.issues.chromium.org/u/1/issues/503987489}{503987489} & \cnf & \no & \no & \no & \no & \no \\
    81  & libaom        & bitstream.c:2475                     & Assertion Fail    & \iss{https://aomedia.issues.chromium.org/issues/471031723}{471031723} & \cnf & \no & \no & \no & \no & \no \\
    82  & protobuf      & coded\_stream.cc:243                 & Null Pointer      & \iss{https://github.com/protocolbuffers/protobuf/issues/26546}{26546} & \cnf & \no & \no & \no & \no & \no \\
    83  & protobuf      & coded\_stream.cc:735                 & Segment Violation & \iss{https://github.com/protocolbuffers/protobuf/issues/27099}{27099} & \cnf & \no & \no & \no & \no & \no \\
    84  & openssl       & ssl\_ciph.c:1241                     & Segment Violation & \iss{https://github.com/openssl/openssl/issues/30919}{30919} & \cnf & \no & \no & \no & \no & \no \\
    85  & openssl       & obj\_lib.c:62                        & Null Pointer      & \iss{https://github.com/openssl/openssl/issues/30922}{30922} & \cnf & \no & \no & \no & \no & \no \\
    86  & openssl       & dh\_kmgmt.c:89                       & Null Pointer      & \iss{https://github.com/openssl/openssl/issues/30921}{30921} & \cnf & \no & \no & \no & \no & \no \\
    87  & openssl       & md32\_common.h:158                   & Type Mismatch     & \iss{https://github.com/openssl/openssl/issues/29615}{29615} & \cnf & \no & \no & \no & \no & \no \\
    88  & openssl       & bsearch.c:28                         & Type Mismatch     & \iss{https://github.com/openssl/openssl/issues/29617}{29617} & \cnf & \no & \no & \no & \no & \no \\
    89  & openssl       & stack.c:443                          & Type Mismatch     & \iss{https://github.com/openssl/openssl/issues/29616}{29616} & \cnf & \no & \no & \no & \no & \no \\
    90  & opencv        & count\_non\_zero.dispatch.cpp:148    & Type Mismatch     & \iss{https://github.com/opencv/opencv/issues/28580}{28580} & \cnf & \no & \no & \no & \no & \no \\
    91  & opencv        & intrin\_sse.hpp:3080                 & Memory Alignment  & \iss{https://github.com/opencv/opencv/issues/28597}{28597} & \cnf & \no & \no & \no & \no & \no \\
    92  & opencv        & drawing.cpp:1121                     & Integer Overflow  & \iss{https://github.com/opencv/opencv/issues/28598}{28598} & \cnf & \no & \no & \no & \no & \no \\
    93  & opencv        & mathfuncs.cpp:141                    & Integer Overflow  & \iss{https://github.com/opencv/opencv/issues/28926}{28926} & \cnf & \no & \no & \no & \no & \no \\
    94  & opencv        & minmax.cpp:1181                      & Type Mismatch     & \iss{https://github.com/opencv/opencv/issues/28928}{28928} & \cnf & \no & \no & \no & \no & \no \\
    95  & opencv        & types.hpp:1245                       & Integer Overflow  & \iss{https://github.com/opencv/opencv/issues/28930}{28930} & \cnf & \no & \no & \no & \no & \no \\
    96  & opencv        & drawing.cpp:1662                     & Integer Overflow  & \iss{https://github.com/opencv/opencv/issues/28940}{28940} & \cnf & \no & \no & \no & \no & \no \\
    97  & opencv        & tree.cpp:522                         & Buffer Overflow   & \iss{https://github.com/opencv/opencv/issues/28941}{28941} & \cnf & \no & \no & \no & \no & \no \\
    98  & opencv        & cap\_mjpeg\_encoder.cpp:178          & Buffer Overflow   & \iss{https://github.com/opencv/opencv/issues/29112}{29112} & \cnf & \no & \no & \no & \no & \no \\
    99  & opencv        & grfmt\_sunras.cpp:72                 & Type Mismatch     & \iss{https://github.com/opencv/opencv/issues/29150}{29150} & \cnf & \no & \no & \no & \no & \no \\
    100 & opencv        & parse\_context.h:725                 & Null Pointer      & \iss{https://github.com/opencv/opencv/issues/29152}{29152} & \cnf & \no & \no & \no & \no & \no \\
    101 & opencv        & copy.cpp:941                         & Integer Overflow  & \iss{https://github.com/opencv/opencv/issues/29232}{29232} & \cnf & \no & \no & \no & \no & \no \\
    102 & opencv        & brisk.cpp:2037                       & Integer Overflow  & \iss{https://github.com/opencv/opencv/issues/29239}{29239} & \cnf & \no & \no & \no & \no & \no \\
\end{longtable}
\endgroup
\twocolumn

\fi

\end{document}